\documentclass[fleqn,usenatbib]{mnras}

\usepackage[T1]{fontenc}
\usepackage{comment}
\DeclareRobustCommand{\VAN}[3]{#2}
\let\VANthebibliography\thebibliography
\def\thebibliography{\DeclareRobustCommand{\VAN}[3]{##3}\VANthebibliography}
\DeclareUnicodeCharacter{2212}{-} 

\usepackage{graphicx}
\usepackage{amsmath}
\usepackage{amssymb}
\usepackage{subfigure}
\usepackage{float}
\usepackage{hyperref}
\usepackage{array}
\usepackage{comment}
\usepackage{caption}
\usepackage{xcolor}
\usepackage{scalerel}
\usepackage{tabularx}

\usepackage{float}
\floatstyle{plaintop}
\restylefloat{table}
\usepackage{xcolor}
\usepackage{tikz}
\definecolor{lime}{HTML}{A6CE39}
\DeclareRobustCommand{\orcidicon}{\hspace{-3mm}
  \begin{tikzpicture}
    \draw[lime, fill=lime] (0,0) 
    circle [radius=0.12] 
    node[white] {\hspace{0.1mm}{\fontfamily{qag}\selectfont \tiny ID}};
    \draw[white, fill=white] (-0.07,0.1) 
    circle [radius=0.01];
  \end{tikzpicture}
  \hspace{-5mm}
}

\foreach \x in {A, ..., Z}{\expandafter\xdef\csname orcid\x\endcsname{\noexpand\href{https://orcid.org/\csname orcidauthor\x\endcsname}
    {\noexpand\orcidicon}}
}

\title[Proximity zones from XQR-30]{New quasar proximity zone size measurements at $z\sim 6$ using the enlarged XQR-30 sample}

\newcommand{\msun}{\mathrm{M}_\odot}
\newcommand{\lya}{Ly$\alpha$}
\newcommand{\lyb}{Ly$\beta$}

\newcommand{\HI}{H\,I}

\newcommand{\HeII}{He\,II}
\newcommand{\HeIII}{He\,III}

\newcommand{\MgII}{Mg\,{\small II}}
\newcommand{\CII}{[C\,{\small II}]}

\newcommand{\NV}{N\thinspace{V}}

\newcommand{\nH}{n_\mathrm{H}}

\newcommand{\rp}{R_\mathrm{p}}

\newcommand{\tq}{t_\mathrm{q}}

\renewcommand{\theenumi}{\textbf{\Alph{enumi}.}}
\renewcommand{\theenumii}{\textbf{\alph{enumii}.}}

\author[Satyavolu et al.]{{Sindhu Satyavolu$^{1}$\thanks{E-mail: sindhu@theory.tifr.res.in} \ \orcidA{}\, ,
    Anna-Christina Eilers$^2$ \ \orcidB{}\, , 
    Girish Kulkarni$^{1}$ \ \orcidC{}\, ,
    Emma Ryan-Weber$^{3,4}$ \ \orcidG{}\, ,
  }
  \newauthor{Rebecca L.~Davies$^{3,4}$ \ \orcidF{}\, ,
  	George D.~Becker$^5$ \ \orcidD{}\, ,
  	Sarah E.~I.~Bosman$^{6,7}$ \ \orcidE{}\, ,
    Bradley Greig$^{4,8}$ \ \orcidM{}\, , 
  }
  \newauthor{Chiara Mazzucchelli$^{9}$ \ \orcidQ{}\, ,
  	Eduardo Ba\~{n}ados$^6$ \ \orcidH{}\, ,
    Manuela Bischetti$^{10,11}$ \ \orcidI{}\, ,
    Valentina D'Odorico$^{10,11,12}$ \ \orcidJ{}\, , 
  }
  \newauthor{Xiaohui Fan$^{13}$ \ \orcidK{}\, ,
  	Emanuele Paolo~Farina$^{14}$ \ \orcidL{}\, ,
    Martin G.~Haehnelt$^{15,16}$ \ \orcidN{}\, , 
    Laura C.~Keating$^{17}$ \ \orcidO{}\, , 
  } 
  \newauthor{Samuel Lai$^{18}$ \ \orcidP{}\, ,
    Fabian Walter$^6$ \ \orcidR{}\,}\\
  $^1$Tata Institute of Fundamental Research, Homi Bhabha Road, Mumbai 400005, India\\
  $^2$MIT Kavli Institute for Astrophysics and Space Research, 77 Massachusetts Avenue, Cambridge, MA 02139, USA\\
    $^3$Centre for Astrophysics and Supercomputing, Swinburne University of Technology, Hawthorn, Victoria 3122, Australia\\
  $^4$ARC Centre of Excellence for All Sky Astrophysics in 3 Dimensions (ASTRO 3D), Australia\\
    $^5$Department of Physics \& Astronomy, University of California, Riverside, CA 92521, USA\\
  $^6$Max Planck Institute for Astronomy, K\"{o}nigstuhl 17, D-69117 Heidelberg, Germany\\
  $^{7}$Institute for Theoretical Physics, Heidelberg University, Philosophenweg 12, D-69120, Heidelberg, Germany\\
  $^{8}$School of Physics, University of Melbourne, Parkville, VIC 3010, Australia\\
    $^{9}${Instituto de Estudios Astrof\'{\i}sicos, Facultad de Ingenier\'{\i}a y Ciencias, Universidad Diego Portales, Avenida Ejercito Libertador 441, Santiago, Chile}\\
  $^{10}$INAF-Osservatorio Astronomico di Trieste, Via Tiepolo 11, I-34143 Trieste, Italy\\
  $^{11}$IFPU-Institute for Fundamental Physics of the Universe, via Beirut 2, I-34151 Trieste, Italy\\
  $^{12}$Scuola Normale Superiore, Piazza dei Cavalieri 7, I-56126 Pisa, Italy\\
  $^{13}$Steward Observatory, University of Arizona, 933 N Cherry Avenue, Tucson, AZ 85721, USA\\
  $^{14}$Gemini Observatory, NSF’s NOIRLab, 670 N A’ohoku Place, Hilo, Hawai'i 96720, USA\\
  $^{15}$Institute of Astronomy, University of Cambridge, Madingley Road, Cambridge CB3 0HA, UK \\
  $^{16}$Kavli Institute of Cosmology, University of Cambridge, Madingley Road, Cambridge CB3 0HA, UK\\
  $^{17}$Institute for Astronomy, University of Edinburgh, Blackford Hill, Edinburgh, EH9 3HJ, UK\\
  $^{18}$Research School of Astronomy and Astrophysics, Australian National University, Canberra, ACT 2611, Australia\\
}

\usepackage{xcolor}
\definecolor{notecolor}{rgb}{0.8,0,0}
\definecolor{color2}{rgb}{0.8,0.0,0.8}

\date{Accepted ---. Received ---; in original form ---}

\pubyear{2023}
\usepackage{newtxtext,newtxmath}
\begin{document}
\label{firstpage}
\pagerange{\pageref{firstpage}--\pageref{lastpage}}
\maketitle 

\begin{abstract}
Proximity zones of high-redshift quasars are unique probes of their central supermassive black holes as well as the intergalactic medium in the last stages of reionization.  We present 22 new measurements of proximity zones of quasars with redshifts between 5.8 and 6.6, using the enlarged XQR-30 sample of high-resolution, high-SNR quasar spectra.  The quasars in our sample have UV magnitudes of $M_{1450}\sim -27$ and black hole masses of $10^9$--$10^{10}$~M$_\odot$.  Our inferred proximity zone sizes are 2--7~physical Mpc, with a typical uncertainty of less than 0.5~physical Mpc, which, for the first time, also includes uncertainty in the quasar continuum.  We find that the correlation between proximity zone sizes and the quasar redshift, luminosity, or black hole mass,  indicates a large diversity of quasar lifetimes. Two of our proximity zone sizes are exceptionally small.  The spectrum of one of these quasars, with $z=6.02$, displays, unusually for this redshift, damping wing absorption without any detectable metal lines, which could potentially originate from the IGM. The other quasar has a high-ionization absorber $\sim$0.5~pMpc from the edge of the proximity zone. This work increases the number of proximity zone measurements available in the last stages of cosmic reionization to 87.  This data will lead to better constraints on quasar lifetimes and obscuration fractions at high redshift, which in turn will help probe the seed mass and formation redshift of supermassive black holes. 
\end{abstract}


\section{Introduction}
\label{sec:intro}

Proximity zones of quasars are unique probes of the growth of supermassive black holes (SMBHs) as well as the intergalactic medium (IGM) between the quasar and us (see \citealt{2022arXiv221206907F} for a review). At redshifts $z\gtrsim6$, quasar proximity zones represent the only regions where there is non-negligible transmission blueward of the quasar's rest-frame \lya\ emission. This can be attributed to the presence of ionizing radiation from the quasar, which carves out a region of ionized hydrogen around itself. Theoretically, the size of such an ionizing bubble depends on the quasar's UV luminosity, its lifetime\footnote{We define the quasar lifetime, $t_\mathrm{q}$ as the time the quasar has spent in the most recent active phase.  In so-called lightbulb models that assume constant quasar lightcurves, $t_\mathrm{q}$ is equal to the total duration for which the quasar, with its constituent black hole, has existed.}, and the amount of neutral hydrogen in the IGM around the quasar. Assuming a uniform gas density around the quasar and spherically symmetric emission at a constant rate, the radius of the bubble, before ionisation equilibrium is reached, is \citep{2007MNRAS.374..493B}
\begin{multline}
  R_{\text{ion}} = 21.2~\mathrm{pMpc}\left(\frac{\dot{N}}{10^{57}\mathrm{s}^{-1}}\right)^{1/3} \left(\frac{t_{\text{q}}}{1~\mathrm{Myr}}\right)^{1/3}  \\ \times \left(\frac{n_{\text{H}}}{7\times10^{-5}~\mathrm{cm^{-3}}}\right)^{-1/3}\left(\frac{x_{\text{HI}}}{10^{-4}}\right)^{-1/3},
  \label{eq:rion}
\end{multline}
where $\dot{N}$ is the number of ionizing photons emitted by the quasar per unit time, $t_{\mathrm{q}}$ is the quasar lifetime or the duration for which the quasar has been emitting this ionizing radiation, and $\nH$ and $x_{\mathrm{HI}}$ are, respectively, the hydrogen density and neutral hydrogen fraction in the IGM around the quasar. 

The size  $R_{\text{ion}}$ of the ionizing bubble around the quasar might not be directly measurable from the quasar spectrum, as the \lya\ transmission becomes insensitive to values of neutral hydrogen fraction above $10^{-4}$ due to saturated absorption.  However, one can define proximity zones that can serve as proxies for the ionized bubbles.  The sizes of proximity zones thus defined have as a result been used to study quasar lifetimes and neutral fraction of the IGM around the quasar. Conventionally, proximity zones are defined as the region blueward of the quasar's \lya\ emission until where the continuum-normalised flux, smoothed by a 20 \AA\ boxcar filter in the observed frame, first drops below 10\% \citep{2006AJ....132..117F}. The size of ionized region $R_{\text{ion}}$ can range up to few tens of proper Mpc for typical values of the quasar and IGM parameters as shown in Equation~(\ref{eq:rion}).  The proximity zone size, $\rp$, however, is limited by the absorption in the IGM and is relatively smaller with typical values of up to a few proper Mpc \citep{2007MNRAS.374..493B}. 

Proximity zone sizes have so far been measured in 75 quasars between redshifts 5.7 and 7.5. \citet{2006AJ....132..117F} were the first to define and measure proximity zones for 16 SDSS $z\sim 6$ quasars. They also defined the luminosity-scaled proximity zones, where the measured proximity zones were corrected to the values that would be measured if all quasars were at a magnitude of $M_{1450}=-27.0$. They found that the luminosity-scaled proximity zone sizes decrease with increasing redshift, and attributed the decline to the evolution of the neutral hydrogen fraction in the IGM at those redshifts.  \citet{2007MNRAS.381L..35B} measured the proximity zone sizes for four SDSS quasars in both \lya\ and \lyb\ forests and suggested that for a large enough sample, their ratio could be used to estimate the volume-averaged neutral fraction.  Following the definition given by \citet{2006AJ....132..117F}, proximity zones for quasars with redshifts $z>$ 5.7 have since been measured by \citet{2007AJ....134.2435W}, \citet{2009A&A...505...97M}, \citet{2010AJ....140..546W}, \citet{2010ApJ...714..834C}, \citet{2011Natur.474..616M}, \citet{2015ApJ...801L..11V}, \citet{2015MNRAS.454.3952R}, \citet{2017ApJ...840...24E}, \citet{2017MNRAS.468.4702R}, \citet{2017ApJ...849...91M}, \citet{2018Natur.553..473B}, \citet{2020ApJ...900...37E}, \citet{2020ApJ...903...60I}, and \citet{2021ApJ...909...80B}.  The highest redshift quasar for which the proximity zone size has been measured is the redshift 7.54 quasar ULAS~J1342+0928 \citep{2018Natur.553..473B}, with a proximity zone size of 1.3~pMpc. The quasars at $z=7.085$ and 7.54 have proximity zone sizes that are three times smaller than the typical values at redshift $z\sim6$. This is because these quasars show damped Ly$\alpha$ absorption by the intergalactic hydrogen.  All of these proximity zone size measurements use similar methods, although they often differ in data quality and some procedural details.  For instance, all measurements exclude broad-absorption-line (BAL) quasars, as the outflow-induced broad absorption lines in these objects can bias the proximity zone size measurement \citep{2020ApJ...900...37E}.  The quasar continuum estimation methods are different in each of the measurements, but while this could lead to differences in the reported proximity zone sizes, \citet{2017ApJ...840...24E} found that in practice the differences are negligible.  

Interpretation of these proximity zone size measurements has led to interesting constraints on the properties of quasars and the IGM.  \citet{2007AJ....134.2435W} estimated luminosity-scaled proximity zone sizes and found them to be relatively large (6.4 and 10.8 pMpc). Following \citet{2007MNRAS.381L..35B}, they concluded that these quasars must be in an already ionized IGM with a neutral hydrogen fraction less than $0.3$ at redshifts 6.1 and 6.43 respectively. \citet{2017ApJ...840...24E} measured proximity zones of 30 quasars between $5.7 \lesssim z \lesssim 6.5 $ and found a much shallower evolution of the luminosity-scaled proximity zone size as a function of redshift, unlike the previous measurements.  They found that this evolution is independent of the IGM around the quasar, suggesting that contrary to previous analyses, the proximity zone size is set by the quasar properties and is relatively insensitive to the neutral hydrogen fraction of the IGM. \citet{2017ApJ...849...91M} and \citet{2020ApJ...903...60I} also found a shallow evolution of proximity zone sizes with redshift. 

\citet{2017ApJ...840...24E} also discovered 3 quasars with proximity zone sizes $< 1$~pMpc.  After confirming that there is no truncation of the proximity zone size due to proximate absorbers or patchy neutral hydrogen islands, they concluded that these quasars must be young with lifetimes $\tq < 10^5$ yr. Such small proximity zones were also found by \citet{2017MNRAS.468.4702R}, who measured proximity zones for four quasars. Two of their quasars showed small luminosity-corrected proximity zones, which they suggest could imply that the quasar is young with $<10^7$--$10^8$~yr age, or that they are located in a region where the average hydrogen neutral density is a factor of 10 higher.  \cite{2020ApJ...900...37E} pre-selected and measured proximity zone sizes for 13 quasars, including two quasars from \citet{2017MNRAS.468.4702R} and one from  \citet{2017ApJ...840...24E}, between $5.8 < z < 6.5$, that were likely to be young after ruling out spurious truncation of proximity zones. They conclude that 5 of their quasars are likely very young quasars with lifetimes $<10^5$ yr. Such short quasar lifetimes have been found to be hard to reconcile with the estimates of the central supermassive black hole masses \citep{2019ApJ...884L..19D,2021ApJ...917...38E}.  Overall, the picture that emerges is that supermassive black holes spend a long time growing in an obscured phase \citep{2023MNRAS.521.3108S} or undergo radiatively inefficient accretion at super/hyper-Eddington rates \citep{2019ApJ...884L..19D, 2021ApJ...917...38E}.  Increasing the sample size of proximity zone studies may therefore enable us to tighten the constraints on black hole growth.

In this paper, we add 22 measurements to the above set of proximity zone size measurements using the XQR-30 sample. This is one of the largest set of proximity zone measurements based on homogeneous, high quality quasar spectra. We use the traditional definition of the proximity zone given by \citet{2006AJ....132..117F}, and examine how the resultant proximity zone sizes correlate with the quasar luminosity, redshift, and black hole mass. We describe our quasar sample and our procedure for measuring the proximity zones in Section~\ref{sec:methods}.  Section~\ref{sec:results} presents our results and discussion. We end with a summary in Section~\ref{sec:conclusions}.  Our measurements as well as theoretical models assume $\Omega_\mathrm{b}=0.0482$, $\Omega_\mathrm{m}=0.308$, $\Omega_\Lambda=0.692$, $h=0.678$, $n_\mathrm{s}=0.961$, $\sigma_8=0.829$, and $Y_\mathrm{He}=0.24$ \citep{2014A&A...571A..16P}.  

\section{Methods}
\label{sec:methods}

\begin{figure*}
  \begin{center}
    \includegraphics[width=0.9\textwidth]{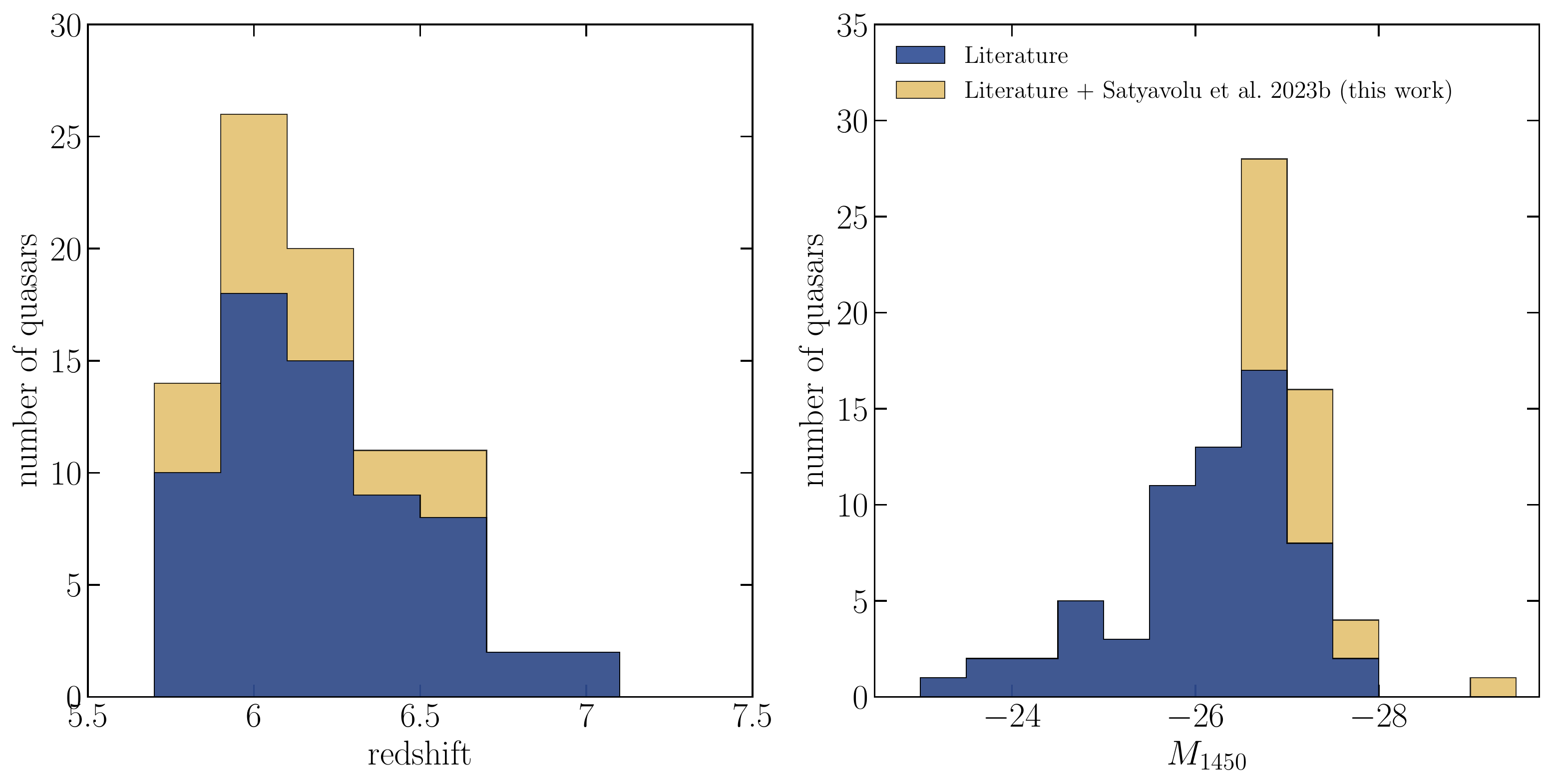}
    \caption{Distribution of quasar redshifts (left panel) and UV magnitudes (right panel) for the 22 quasars studied in this paper.  The blue histograms show the distributions for the 65 quasars for which proximity zones have been measured previously, as discussed in Section~\ref{sec:intro}, after excluding the 10 quasars for which we have updated the proximity zone size measurements in this work. The yellow histograms show the distributions for all 87 quasars for which proximity zones sizes are now available, including the 22 that have been measured in this work. \label{fig:zdatahist}}
  \end{center}
\end{figure*}

  \begin{table*}
    \begin{center}
      \begin{tabular} {p{0.02\textwidth} p{0.15\textwidth} p{0.15\textwidth}  p{0.05\textwidth} p{0.1\textwidth} p{0.23\textwidth}p{0.19\textwidth}}
        \hline 
        & Object & $z_\mathrm{qso}$ & Line & $M_{1450}$ & Ref.\ (redshift) & Ref.\ (magnitude)\\
        \hline
        1 & J0408-5632  & $6.033^{\scaleto{+0.0107}{4pt}}_{\scaleto{-0.006}{4pt}}$& \MgII & $-26.56$ &Bischetti et al. (\citeyear{2022Natur.605..244B})& Reed et al. (\citeyear{2017MNRAS.468.4702R}) \\ 
        2 & PSOJ029-29 & $5.976^{+\scaleto{+0.0106}{4pt}}_{\scaleto{-0.006}{4pt}}$ & \MgII & $-27.32$& Bischetti et al. (\citeyear{2022Natur.605..244B}) &Ba{\~n}ados et al. (\citeyear{2016ApJS..227...11B}) \\ 
        3 & ATLASJ029-36 & $ 6.013^{\scaleto{+0.0106}{4pt}}_{\scaleto{-0.006}{4pt}}$ & \MgII &  $-27.00$& Bischetti et al. (\citeyear{2022Natur.605..244B})& Ba{\~n}ados et al. (\citeyear{2016ApJS..227...11B})\\
        4 & VDESJ0224-4711 & $6.525^{\scaleto{+0.0114}{4pt}}_{\scaleto{-0.0064}{4pt}}$  & \MgII & $-26.98$&Bischetti et al. (\citeyear{2022Natur.605..244B})& Reed et al. (\citeyear{2017MNRAS.468.4702R})\\
        5 & PSOJ060+24 & $6.17^{\scaleto{+0.0109}{4pt}}_{\scaleto{-0.0061}{4pt}}$  & \MgII & $-26.95$ &Bischetti et al. (\citeyear{2022Natur.605..244B}) &Ba{\~n}ados et al. (\citeyear{2016ApJS..227...11B})\\
        6 & PSOJ108+08 & 5.9647$\pm$0.0023  &  \CII & $-27.59$ &Bosman et al. (in prep.)& Ba{\~n}ados et al. (\citeyear{2016ApJS..227...11B})\\
        7 & SDSSJ0842+1218 & 6.0754$\pm$ 0.0024 &  \CII &$-26.91$& Schindler et al. (\citeyear{2020ApJ...905...51S})&Ba{\~n}ados et al. (\citeyear{2016ApJS..227...11B})\\
        8 & PSOJ158-14 & 6.0687$\pm$   0.0024 &  \CII&$-27.32$& Bosman et al. (in prep.)&Ba{\~n}ados et al. (\citeyear{2023ApJS..265...29B}) \\
        9 & PSOJ183-12 &  $5.893^{\scaleto{+0.0105}{4pt}}_{\scaleto{-0.0059}{4pt}}$  & \MgII & $-27.49$&D'Odorico et al. (\citeyear{xqr30general})&Ba{\~n}ados et al. (\citeyear{2016ApJS..227...11B})\\
        10& PSOJ217-16 &  6.1466$\pm$ 0.0024 &  \CII & $-26.94$&Bosman et al. (in prep.)& Ba{\~n}ados et al. (\citeyear{2016ApJS..227...11B})\\
        11& PSOJ242-12 & 5.8468$\pm$ 0.0023 &  \CII & $-26.92$&Bosman et al. (in prep.)&Ba{\~n}ados et al. (\citeyear{2016ApJS..227...11B})\\
        12& PSOJ308-27 & $5.799^{\scaleto{+0.0103}{4pt}}_{\scaleto{-0.0058}{4pt}}$  & \MgII & $-26.78$ &D'Odorico et al. (\citeyear{xqr30general})& Ba{\~n}ados et al. (\citeyear{2016ApJS..227...11B})\\
        13& PSOJ323+12 & 6.5872$\pm$0.0025 & \CII & $-27.07$ &Schindler et al. (\citeyear{2020ApJ...905...51S})&Mazzucchelli et al. (\citeyear{2017ApJ...849...91M})\\
        14& PSOJ359-06 & 6.1719$\pm$ 0.0024 & \CII & $-26.79$ &Schindler et al. (\citeyear{2020ApJ...905...51S})& Ba{\~n}ados et al. (\citeyear{2016ApJS..227...11B})\\
        15& SDSSJ0927+2001 & 5.7722$\pm$0.0023 & CO &$-26.76$&Wang et al. (\citeyear{2010ApJ...714..699W})& Ba{\~n}ados et al. (\citeyear{2016ApJS..227...11B})\\
        16& SDSSJ0818+1722 & $5.967^{\scaleto{+0.0105}{4pt}}_{\scaleto{-0.0059}{4pt}}$&\MgII&$-27.52$&D'Odorico et al. (\citeyear{xqr30general})& Ba{\~n}ados et al. (\citeyear{2016ApJS..227...11B})\\
        17& SDSSJ1306+0356 & 6.033$\pm$0.0023&\CII&$-27.15$&Decarli et al. (\citeyear{2018ApJ...854...97D})& Nanni et al. (\citeyear{Nannietal})\\
        18& ULASJ1319+0950 & 6.1347$\pm$0.0024 &\CII&$-27.05$& Venemans et al. (\citeyear{2020ApJ...904..130V})& Ba{\~n}ados et al. (\citeyear{2016ApJS..227...11B})\\
        19& SDSSJ1030+0524 &$ 6.309^{\scaleto{+0.0111}{4pt}}_{\scaleto{-0.0062}{4pt}}$ &\MgII&$-26.99$&Jiang et al. (\citeyear{2007AJ....134.1150J})& Ba{\~n}ados et al. (\citeyear{2016ApJS..227...11B})\\
        20& SDSSJ0100+2802&6.3269$\pm$0.0024&\CII&$-29.14$&Wang et al. (\citeyear{2016MNRAS.460.2143W})&Ba{\~n}ados et al. (\citeyear{2016ApJS..227...11B})\\
        21& ATLASJ025-33&6.3373$\pm$0.0024 &\CII&$-27.50$&Decarli et al. (\citeyear{2018ApJ...854...97D})& Carnall et al. (\citeyear{2015MNRAS.451L..16C})\\
        22& PSOJ036+03 & 6.5405$\pm$0.0025 &\CII&$-27.33$&Venemans et al. (\citeyear{2020ApJ...904..130V})& Ba{\~n}ados et al. (\citeyear{2016ApJS..227...11B})\\
        \hline
      \end{tabular}
      \caption{Properties of the 22 quasars studied in this paper.  The columns show the serial number, quasar name, quasar redshift with the total 1$\sigma$ uncertainty, the emission line used for determining the quasar redshift, quasar absolute UV magnitude at 1450~\AA, and references for the quasar redshift and magnitude. \label{tab:qso-props} }
    \end{center}
  \end{table*}

XQR-30 is an European Southern Observatory (ESO) Large Programme (ID:~1103.A-0817, P.I.\ V.~D'Odorico) that targeted 30 quasars with redshifts between 5.8 and 6.6 using VLT/XSHOOTER \citep{2011A&A...536A.105V} to obtain high-resolution, high-SNR rest-frame UV spectra.  The target quasars are some of the brightest quasars known in the southern hemisphere in this redshift range \citep{xqr30general}. The spectra were taken with slit widths of 0.9~arcsec and 0.6~arcsec, nominal resolution $R\sim 8900$ and $8100$, and median resolution of $R\sim 11400$ and $9800$ in the visible (VIS) and near-infrared (NIR) arms of XSHOOTER, with pixel size of 10~km$/$s in both arms (Resolution, however, is not a deciding factor in our work, since we smooth the spectra by a 20~\AA\ boxcar for obtaining the proximity zone size). The observing time on target ranged from 4~h to 11~h.  The median SNR per pixel in the rest-frame 1600--1700~\AA\ wavelength range is between 25 and 160 for spectra rebinned to 50~km/s.  Data reduction, which includes optimal sky subtraction, telluric absorption correction, optimal extraction and direct combination of exposures, was done using a custom IDL pipeline developed for the XQ-100 survey \citep{2019ApJ...883..163B} with minor improvements, mainly for the NIR arm.  Further details about data reduction will be discussed by \citet{xqr30general}.  We also include 12 archival VLT/XSHOOTER spectra in our sample, that, together with the 30 XQR-30 quasars, form the enlarged XQR-30 sample.  These have similar redshifts, magnitudes, SNR, and comparable spectral resolution as the XQR-30 sample. The data reduction for these additional quasars was done with the same pipeline that was used for the XQR-30 sample. The full sample is described in \citet{2022MNRAS.514...55B} and will also be discussed in \citet{xqr30general}. 

Of the 42 quasars in the enlarged XQR-30 sample, we use 22 in this study.  We exclude 12 quasars that show strong broad absorption lines \citep[BALs;][]{2022Natur.605..244B} and 7 quasars with proximate damped \lya\ systems \citep[pDLAs;][]{2023MNRAS.521..289D,2019ApJ...885...59B}. We exclude BAL quasars because their proximity zones may be affected by unseen strong \NV\ associated absorption.
Proximate damped Lyman-$\alpha$ systems are absorption systems with neutral hydrogen column density $N_\mathrm{HI}>2\times10^{20}\mathrm{cm}^{-2}$ at a velocity separation $\Delta v<3000 \,\mathrm{km\,s^{-1}}$ from the quasar \citep{2008ApJ...675.1002P}.  pDLAs can prematurely truncate the quasar flux, leading to spuriously small proximity zones. We exclude all quasars with pDLAs at a velocity separation $\Delta v<5000 \,\mathrm{km\,s^{-1}}$ from the quasar, that have been identified by the presence of neutral oxygen tracing the neutral hydrogen or by their associated ionised absorbers \citep[Sodini et al. in preparation]{2023MNRAS.521..289D}. They are also not modelled in our simulations, making them not suitable for comparison. We also exclude the heavily reddened quasar J1535+1943, which is most likely obscured \citep{2021ApJ...923..262Y}. The large error on the systemic redshift of this quasar makes a reliable measurement of its proximity zone size difficult.

We obtain the normalised transmitted flux by fitting continuum spectra redward of the quasar's \lya\ line using the log-PCA approach of \citet{2018ApJ...864..143D}, as described in \citet{2022ApJ...931...29C} and \cite{2022MNRAS.514...55B}.  This method improves upon the original PCA-based continuum fitting introduced by \citet{2005astro.ph..3248S} and \citet{2011A&A...530A..50P}. We note however that the choice of continuum fitting method has been found to have a negligible impact on the proximity zone size measurement \citep{2017ApJ...840...24E}.

\subsection{Quasar redshifts and magnitudes}

Table~\ref{tab:qso-props} summarises the redshifts and magnitudes of the 22 quasars in our study. Accurately measuring the redshifts of these quasars is difficult but also necessary for accurate estimates of the proximity zone sizes.  For 13 of the 22 quasars, we use redshifts determined from the emission lines due to the transitions of CO or \CII\ from the host galaxy  \citep[][Bosman et al.\ in preparation]{2018ApJ...854...97D,2010ApJ...714..699W}.  We assign an uncertainty to this redshift measurement of $\Delta \varv\sim 100$~km/s, corresponding to blueshift of the emission line from the quasar's systemic rest-frame.  The uncertainty associated with the fit to the emission lines is negligible. For the remaining 9 quasars, we use the redshifts measured from the quasar's Mg~II emission line \citep{xqr30general,2022Natur.605..244B}, with a typical associated uncertainty of $\Delta \varv\sim 391$~km/s \citep{2020ApJ...905...51S}.

The absolute magnitude at 1450 \AA\ ($M_{1450}$) is measured from the apparent magnitude $m_{1450}$, which is obtained by extrapolating the magnitude in the $y_\mathrm{P1}$ or $J$ bands, depending on where contamination due to emission lines is lower, using a power law shape for continuum with spectral index $\alpha=-0.3$ \citep{2016ApJS..227...11B}. The references for absolute magnitudes for each of the quasars are listed in Table~\ref{tab:qso-props}.

Figure~\ref{fig:zdatahist} shows the distribution of redshifts and magnitudes of quasars for which proximity zones have been previously measured (see Section~\ref{sec:intro}) and our addition to this distribution.  Our sample significantly increases the number of proximity zone sizes measured for quasars with redshifts $5.9<z<6.1$ and with magnitudes $-27.5<M_{1450}<-26.5$.

\section{Results}
\label{sec:results}

\begin{figure*}
  \begin{center}
    \includegraphics[width=0.85\textwidth]{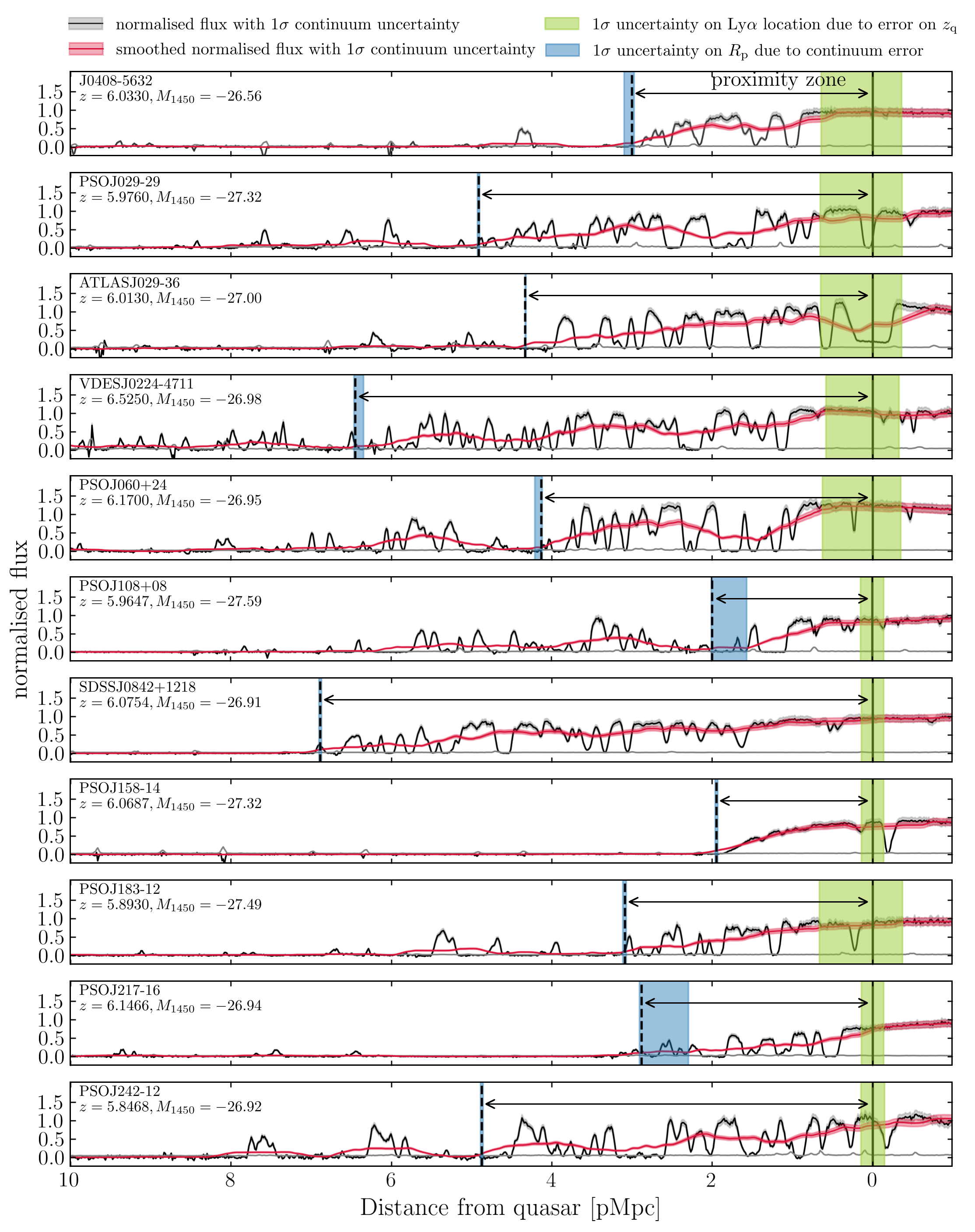}
    \caption{Proximity zones of the quasars in our sample.  The normalised flux obtained by dividing measured flux by continuum, is shown in black. Red curves show the smoothed spectra with shaded region showing the 1$\sigma$ uncertainty in the continuum. Black solid and dotted lines show the quasar location and the extent of proximity zones, respectively. The blue shaded regions show the 1$\sigma$ uncertainty on proximity zone sizes due to continuum uncertainties.  Green shaded regions show redshift errors as the uncertainty on the location of the expected \lya\ emission of the quasar.  \label{fig:fullsample}}
  \end{center}
\end{figure*}

\begin{figure*}
  \begin{center}
  	\hspace*{10mm}
    \includegraphics[width=0.85\textwidth]{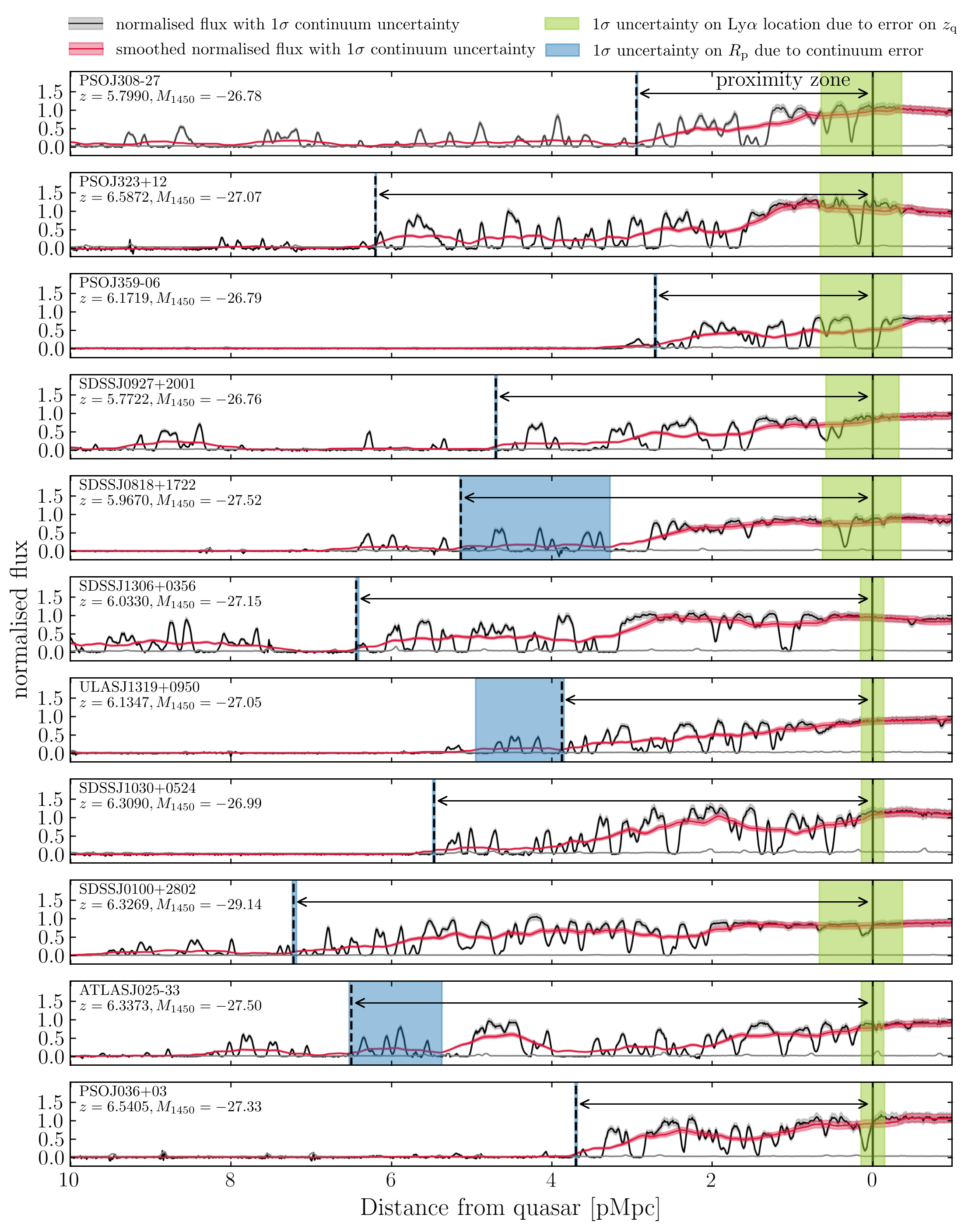}
    \contcaption{}
  \end{center}
\end{figure*}

\subsection{Proximity zone sizes}

\begin{figure}
  \centering
  \includegraphics[width=\columnwidth]{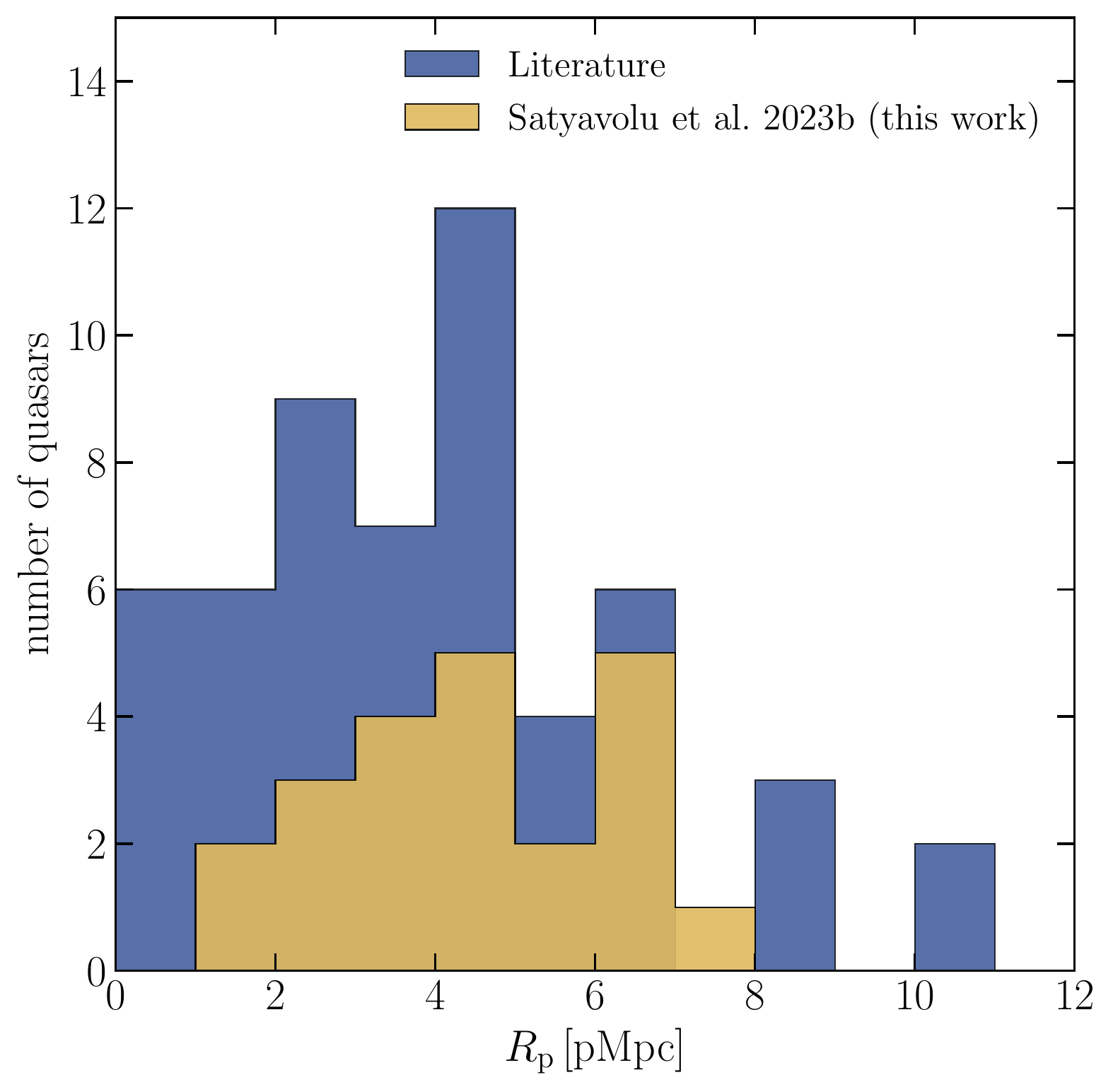}
  \caption{Distribution of proximity zone sizes reported in this work.  The blue histogram shows the distribution of all previously available proximity zone sizes \citep{2010ApJ...714..834C, 2017ApJ...840...24E, 2017ApJ...849...91M, 2017MNRAS.468.4702R, 2018Natur.553..473B, 2020ApJ...900...37E, 2020ApJ...903...60I, 2021ApJ...909...80B}, except those only available as values scaled to a fiducial luminosity, or that have been updated in this paper.  The yellow histogram shows the distribution of the 22 proximity zone sizes presented in this work.}
  \label{fig:rphist}
\end{figure}

To measure the proximity zone sizes of the quasars in our sample, we follow the convention introduced by \citet{2006AJ....132..117F}.  We smooth the continuum-normalised flux of each quasar by a 20~\AA\ boxcar in the observer's frame, and locate the pixel with redshift $z_{\mathrm{GP}}$ at which the smoothed normalised flux first drops below 0.1.  The proximity zone size $R_{\mathrm{p}}$ is then calculated by dividing the comoving line of sight distance between $z_{\mathrm{qso}}$ and $z_{\mathrm{GP}}$ by (1+$z_{\mathrm{qso}}$) to obtain the proper distance.  Figures~\ref{fig:fullsample} shows the resulting proximity zones.  Table~\ref{tab:qso-rp} lists the proximity zone sizes.  Figure~\ref{fig:rphist} shows their distribution.  

Figure~\ref{fig:fullsample} shows the spectra and corresponding proximity zones for all the quasars in our sample.  The red curves show the smoothed spectrum with shaded regions showing $1 \sigma$ spread due to continuum uncertainties.  Instrumental noise on the spectrum is negligible and hence we do not propagate this error onto the proximity zone size.  Following \citet{2017ApJ...840...24E}, the error on the proximity zone size due to redshift uncertainty is calculated as $\Delta \rp= \Delta \varv/H(z)$, where $ \Delta \varv$ is the redshift uncertainty in velocity units.  The quasars in our sample have $\Delta \varv =100$  km/s (for $[\mathrm{C}\, {\textsc{II}}]$  redshifts) which corresponds to an uncertainty of $\Delta \rp \sim 0.14$ pMpc in the proximity zone size at redshift 6. The uncertainty is larger for quasars with $\mathrm{Mg}\, {\textsc{II}}$ redshifts, with a median value of $\Delta \rp \sim 0.5$ pMpc. The continuum errors are computed by measuring the proximity zone sizes of the $1\sigma$ upper and lower bounds of the continuum-normalised flux using the same definition. For most of our quasars, the redshift uncertainty errors dominate over the continuum uncertainty errors on the proximity zone sizes, as shown in Table~\ref{tab:qso-rp}. All previous analyses are thus justified in neglecting the continuum errors. The largest error on $R_{\mathrm{p}}$  due to continuum uncertainties is observed in the archival quasar SDSSJ0818+1722 to be 1.86 proper Mpc ($\sim$~36$\%$ of the measured value), even though the 1$\sigma$ uncertainty on the continuum is not significant. This is because the definition of proximity zone size is such that even though the smoothed flux is quite close to 0.1 due to the uncertainty of the continuum placement, $\rp$ is not defined until the smoothed flux becomes equal to or less than 0.1. Likewise, even though the proximity zones of some quasars are of similar size (e.g PSOJ158-14 and PSOJ108+08), their flux outside the proximity zone size is quite different. In order to better constrain quasar lifetimes based on proximity zone sizes, we will study the use of multiple definitions for proximity zone sizes based on the flux threshold in future work (Satyavolu et al., in preparation).

The total error on the proximity zone sizes of the quasars was obtained by adding the redshift and continuum uncertainty errors in quadrature. All the proximity zone measurements with their errors are shown in Table~\ref{tab:qso-rp}. Out of these, proximity zones were previously measured for ten quasars of our present sample. We have updated proximity zone measurements for the quasars PSOJ060+24, SDSSJ0100+2802, SDSSJ0818+1722, PSOJ036+03 \citep{2017ApJ...840...24E}, PSOJ323+12 \citep{2017ApJ...849...91M},   PSOJ158-14, PSOJ359-06 \citep{2020ApJ...900...37E} and SDSSJ0927+2001, ULASJ1319+0950,  SDSSJ1030+0524  \citep{2020ApJ...903...60I} with the latest redshifts and X-SHOOTER spectra. The newer measurements differ from the older measurements by $\sim 1\%$ to not more than $5\%$. The minor differences are expected to be due to difference in redshifts. One quasar ULASJ1319+0950 is reported to have a proximity zone size of 4.99~pMpc from \citet{2020ApJ...903...60I}. Our updated measurement of 3.87~pMpc is closer to the value of 3.84~pMpc measured by \citet{2017ApJ...840...24E}.

Figure~\ref{fig:rphist} shows the distribution of the proximity zone sizes of the enlarged XQR-30 sample. The largest and smallest proximity zones we measure are 7.22 and 1.95~pMpc, with a median around 5~pMpc. Also shown in blue is the distribution of all previously measured proximity zone sizes, not scaled to a fiducial quasar luminosity and excluding the ten quasars that have been updated in this work. Our proximity zone sizes are consistent with previous measurements, and add to the number of small proximity zone sizes ($\rp<2$~pMpc) measured in the literature recently.

\begin{table*}
  \begin{center}
    \begin{tabular}{p{0.05\textwidth} p{0.16\textwidth} p{0.12\textwidth} p{0.08\textwidth} p{0.09\textwidth} p{0.08\textwidth} p{0.09\textwidth} p{0.08\textwidth} p{0.08\textwidth}}
      \hline
      &&& \multicolumn{2}{p{0.2\linewidth}}{Continuum error ($\Delta\rp$)}&\multicolumn{2}{p{0.2\linewidth}}{\;\;Redshift error ($\Delta\rp$)}&\multicolumn{2}{p{0.2\linewidth}}{\;\;\;Total error ($\Delta\rp$)}\\
      & Object & $\rp$ & Lower $1\sigma$&Upper 1$\sigma$&Lower $1\sigma$&Upper 1$\sigma$&Lower $1\sigma$&Upper 1$\sigma$\\
      & & (pMpc)  & (pMpc) &(pMpc) &(pMpc) &(pMpc) &(pMpc) &(pMpc) \\
      \hline
      1  & J0408-5632     & 3.00    & 0.03 & 0.10  & 0.36 & 0.65 & 0.37 & 0.66 \\
      2  & PSOJ029-29     & 4.91 & 0.01 & 0.01 & 0.37 & 0.66 & 0.37 & 0.66 \\
      3  & ATLASJ029-36   & 4.33 & 0.01 & 0.01    & 0.37 & 0.65 & 0.37 & 0.65 \\
      4  & VDESJ0224-4711 & 6.45 & 0.10  & 0.01 & 0.33 & 0.59 & 0.35 & 0.59 \\
      5  & PSOJ060+24*     & 4.13 & 0.01 & 0.08 & 0.35 & 0.63 & 0.35 & 0.63 \\
      6  & PSOJ108+08     & 1.99 & 0.43 & 0.01 & 0.14 & 0.14 & 0.46 & 0.15 \\
      7  & SDSSJ0842+1218 & 6.89 & 0.01 & 0.01 & 0.14 & 0.14 & 0.14 & 0.14 \\
      8  & PSOJ158-14*     & 1.95 & 0.01 & 0.01 & 0.14 & 0.14 & 0.14 & 0.14 \\
      9  & PSOJ183-12     & 3.09 & 0.01 & 0.03 & 0.38 & 0.67 & 0.38 & 0.67 \\
      10 & PSOJ217-16     & 2.88 & 0.58 & 0.03 & 0.14 & 0.14 & 0.6  & 0.14 \\
      11 & PSOJ242-12     & 4.87 & 0.01 & 0.01 & 0.15 & 0.15 & 0.15 & 0.15 \\
      12 & PSOJ308-27     & 2.95 & 0.01 &  0.01    & 0.38 & 0.68 & 0.38 & 0.68 \\
      13 & PSOJ323+12*     & 6.20  & 0.01 &  0.01    & 0.13 & 0.13 & 0.13 & 0.13 \\
      14 & PSOJ359-06*    & 2.71 & 0.01 & 0.01 & 0.14 & 0.14 & 0.14 & 0.14 \\
      15 & SDSSJ0927+2001* & 4.70  & 0.02 & 0.02 & 0.15 & 0.15 & 0.15 & 0.15 \\
      16 & SDSSJ0818+1722* & 5.13 & 1.86 & 0.01 & 1.43 & 1.43 & 2.35 & 1.43 \\
      17 & SDSSJ1306+0356 & 6.43 & 0.03 &  0.01    & 0.14 & 0.14 & 0.15 & 0.14 \\
      18 & ULASJ1319+0950* & 3.87 & 0.03 & 1.07 & 0.14 & 0.14 & 0.14 & 1.08 \\
      19 & SDSSJ1030+0524* & 5.47 & 0.01 & 0.01 & 0.34 & 0.61 & 0.34 & 0.61 \\
      20 & SDSSJ0100+2802* & 7.22 & 0.04 & 0.01 & 0.13 & 0.13 & 0.14 & 0.13 \\
      21 & ATLASJ025-33   & 6.50  & 1.13 & 0.03 & 0.13 & 0.13 & 1.13 & 0.14 \\
      22 & PSOJ036+03*     & 3.70 & 0.01 & 0.01 & 0.13 & 0.13 & 0.13 & 0.13\\
      \hline
      \multicolumn{9}{p{\linewidth}}{* Previously available measurements that have been updated in this work.}
    \end{tabular}
    \caption{Our proximity zone size measurements.  Columns show the serial number, the name of the quasar, proximity zone size in proper~Mpc with the continuum, redshift and total uncertainties. The minimum error on $\rp$ due to continuum uncertainties is the spatial resolution of the spectra, which is $\sim 0.01$~pMpc. Total error is obtained by adding the continuum and redshift errors in quadrature. \label{tab:qso-rp}}
  \end{center}
\end{table*}

\subsection{Radiative transfer simulations}\label{sec:sim}

Cosmological radiative transfer simulations are necessary to interpret proximity zone size measurements \citep[e.g.,][]{2015MNRAS.454..681K}.  In order to simulate proximity zones, we use the set-up described in our previous work \citep{2023MNRAS.521.3108S}.  We report the essential steps of the procedure here, and direct the reader to that paper for further details.  We combine 3D cosmological hydrodynamical and radiative transfer simulations with a 1D radiative transfer simulation to obtain \lya\ absorption spectra. The underlying simulation volume is generated by post-processing a cosmological hydrodynamical simulation run with P-GADGET3 (modified version of GADGET-2 which is discussed in \citealt{2005MNRAS.364.1105S}) using the radiative transfer code ATON \citep{2008MNRAS.387..295A, 2010ApJ...724..244A} as described by \citet{2019MNRAS.485L..24K}. The box size is 160~cMpc$/h$ box with $2048^3$ gas and dark matter particles. Hydrogen reionization ends at $z=5.3$ in our simulations, with the process half-complete at $z=7$.  This model is consistent with a variety of high-redshift data  \citep{2020A&A...641A...6P,2020MNRAS.491.1736K,2015MNRAS.447.3402B,2017MNRAS.466.4239G,2019MNRAS.484.5094G,2018ApJ...864..142D,2020ApJ...896...23W,2020MNRAS.497..906K,2018MNRAS.479.2564W,2019MNRAS.485.1350W}, and continues to be consistent with newer measurements of the \lya\ forest from the XQR-30 programme \citep{2022MNRAS.514...55B}. The mean free path of hydrogen ionising photons in our simulations at redshift $z\sim6$ is $1\sigma$ larger than the mean free path measured by \citet{2021MNRAS.508.1853B}.  If this difference proves to be correct, the measured proximity zone sizes in our simulated spectra could be systematically larger than true values due to missing structure in the IGM. The highest and lowest mass of halos in our simulations are $4.59\times10^{12}$ and $2.32\times10^8\,\msun$, respectively. The spatial resolution is $\sim 78$~kpc$/h$.  For the proximity zone modelling, we draw sightlines along different directions such that they start on halos, and process these with our 1D radiative transfer algorithm \citep{2023MNRAS.521.3108S} assuming a quasar with a given luminosity and spectrum at the starting point of the sightline. The radiative transfer algorithm computes the ionisation fractions of \HI, \HeII, \HeIII\ and gas temperatures given the initial conditions. The initial ionisation, densities and temperature around the quasars along the 1D skewers are set by our 3D simulations.  The background photoionisation rates are assigned by assuming photoionisation equilibrium with the IGM without the quasar.

We use the \citet{2015MNRAS.449.4204L} model for quasar spectra.  Given the magnitude of the quasar, the specific luminosity at 1450\AA\ can be calculated as 
\begin{equation}
  L_{1450} = 10^{(51.60-M_{1450})/2.5}  \mathrm{erg\, s^{-1} Hz^{-1}}
\end{equation}
The specific luminosity of quasar is then derived by assuming the quasar SED to be a broken power law with a spectral index of 
\begin{equation}
  L_\nu\propto\begin{cases}
  \nu^{-0.61} & \text{if}~\lambda\ge 912~\text{\AA},\\
  \nu^{-1.70} & \text{if}~\lambda<912~\text{\AA}.                
  \end{cases}
\end{equation}
The number of hydrogen-ionizing photons emitted by the quasar per unit time is given by 
\begin{equation}
  \dot{N} = \int_{\nu_{\text{HI}}}^{\infty}\frac{L_{\nu}}{h\nu}d\nu.
\end{equation}
We assume the quasar light curve to be such that the quasar stays on continuously throughout its lifetime (this is known as the `lightbulb' model).   A different lightcurve in which the quasar turns on and off periodically with a duty cycle and episodic lifetime (this is known as the `flickering lightcurve' model) is also useful to consider, particularly for the smallest proximity zones \citep{2023MNRAS.521.3108S}, but we leave the comparison of our measurements to the latter for future work (Satyavolu et al., in preparation). The hydrogen and helium densities are assumed to be constant throughout the computation and equal to those at the redshift of the quasar as we do not expect them to change by much during the quasar lifetimes we consider (up to $\sim 100$~Myr). We combine the neutral hydrogen density and temperature from the output of 1D radiative transfer with peculiar velocities from the 3D simulations to compute the \lya\ optical depth $\tau$ along the line of sight assuming a Voigt absorption profile \citep{2006MNRAS.369.2025T}.  The transmitted flux is calculated as $F = \exp(-\tau)$. We compute the proximity zone size from the model spectra in the same way as the observed spectra: we smooth the flux with a boxcar filter of 20\;\AA\ in the observed-frame and  calculate the proximity zone size as the distance at which the smoothed flux drops below 0.1. 

\subsection{Correlation of proximity zone sizes with quasar luminosity}

\begin{figure}
  \includegraphics[width=\columnwidth]{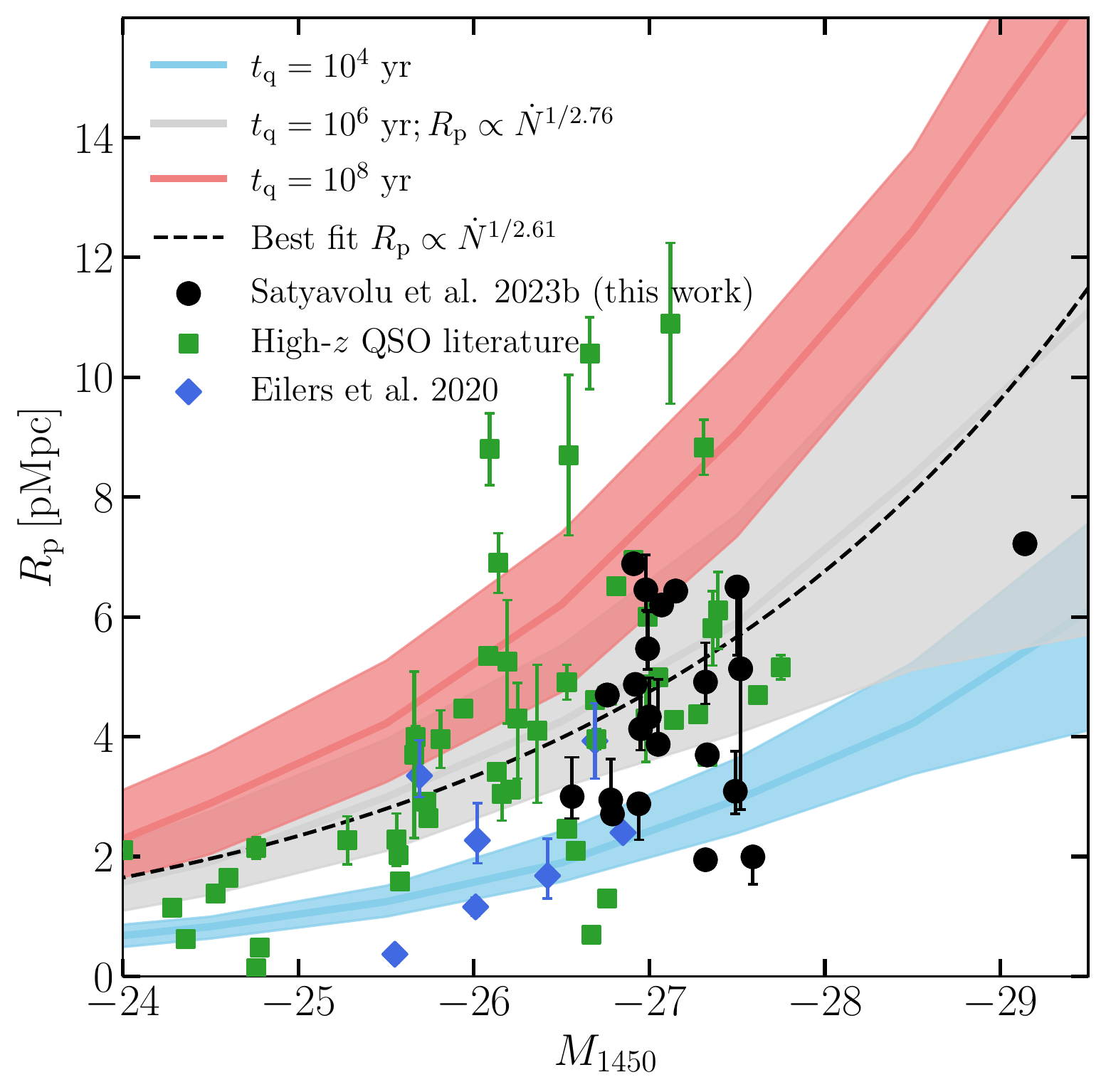}
  \caption{Proximity zone sizes as a function of quasar magnitude. Previous measurements are shown in green. The targeted sample of \citet{2020ApJ...900...37E} is shown in blue. Our measurements are shown in black. The errors on our proximity zone sizes are due to both continuum and redshift uncertainties. The blue, grey and red curves are from our simulations for quasar ages of $10^4$, $10^6$ and $10^8$~yr at a redshift of 5.95. Shaded regions show 68\% scatter across 500 sightlines from our simulations. The black dotted line shows the best fit curve to a relatively homogeneous subset of quasars with $6<z<6.2$, except quasars from the targeted sample of \citet{2020ApJ...900...37E}. \label{fig:rpvmagfit}}
\end{figure}

Figure~\ref{fig:rpvmagfit}  shows the distribution of $\rp$ as a function of quasar magnitude.  It can be seen that although the quasars in our sample have very similar magnitudes, with mostly $-26.5<M_{1450}<-27.5$, the proximity zone size distribution can vary considerably.  The smallest proximity zone is found at a magnitude of $-27.51$ and the largest proximity zone at a magnitude of $-29.14$, both at similar redshifts of 6.06 and 6.32, respectively.  Most of the measured values are consistent with earlier measurements at similar redshifts ($z\sim6$). 

Also shown are the median proximity zone sizes and the $1\sigma$ distribution around the median values from our simulations, for a lightbulb quasar. The median proximity zone size increases with increase in quasar lifetime, as the longer the quasar is active, the farther its ionisation front can travel before reaching the equilibrium value. For brighter quasars, there is also an increase in the spread of the proximity zone size distribution before the quasar lifetime reaches the equilibration timescale. This can be understood as a consequence of the ionisation fronts traveling farther in a small enough time, and encountering neutral hydrogen islands along random directions. For a fainter quasar, the quasar will need more time for its ionisation front to travel farther and encounter such neutral islands. Therefore, fainter quasars see only their immediate surroundings, which are almost uniformly ionised at these redshifts and lifetimes, leading to a narrower spread. The $1\sigma$ spread is also the largest for $\tq\sim 10^6$~yr for similar reasons, as a younger quasar and an older quasar see a mostly ionised medium.
The large proximity zones in our sample are consistent with the models of lightbulb quasars of age $\geq$ 1~Myr.  The smaller proximity zones with $R_{\mathrm{p}}\lesssim 2$~pMpc appear to indicate a young lifetime of $\lesssim 10^4$ yr for a lightbulb quasar at a redshift $z\sim6$.  The fraction of such quasars with small proximity zones is 2 out of 22 or about $\sim 9$ percent in our sample, consistent with the fraction of $5$--$10\%$ estimated by \citet{2020ApJ...900...37E}. We discuss the two smallest $\rp$ values in greater detail in Section~\ref{sec:smallrp} below. 

In order to study the correlation of proximity zone sizes with quasar magnitudes without being influenced by the redshift of the quasars, we obtain a best-fit curve to all measured proximity zone sizes  \citep[excluding the targeted sample of][] {2020ApJ...900...37E} including ours against their magnitudes for quasars with redshifts between $6<z<6.2$ assuming a power law between $\rp$ and $\dot{N}$.  The redshift range was chosen such that the number of quasars for which proximity zone sizes are measured is maximized (see Figure~\ref{fig:zdatahist}).  In a mostly uniform medium, the scaling follows $\rp \propto \dot{N}^{1/3}$  while in a mostly ionised medium, $\rp \propto \dot{N}^{1/2}$ \citep{2007MNRAS.374..493B}. Since at the redshifts of our quasars, the universe is most likely to be partly ionised and partly uniform, one could expect the scaling to fall between $\rp \propto \dot{N}^{1/2}$ and $\rp \propto\dot{N}^{1/3}$, depending on the redshift of the quasar.  Our simulations find an evolution of $\rp \propto \dot{N}^{1/2.76}$, for a quasar lifetime of 1~Myr and redshift 5.95. The best fit to all data within the redshift range $6<z<6.2$ shows an evolution of $\rp \propto \dot{N}^{1/2.61}$, slightly steeper than the scaling inferred from our simulations, but consistent within the expected range for the scaling at this redshift.

\subsection{Correlation of proximity zone sizes with quasar redshift}\label{sec:rpvz}

The evolution of proximity zone sizes as a function of redshift encodes information about reionization \citep{2023MNRAS.521.3108S}.  Models in which reionization ends later cause a 30\% reduction in proximity zone sizes and increase the scatter in their distribution by 10\%, as the growth of ionization fronts is impeded by neutral parts of the IGM.  

\begin{figure}
  \includegraphics[width=\columnwidth]{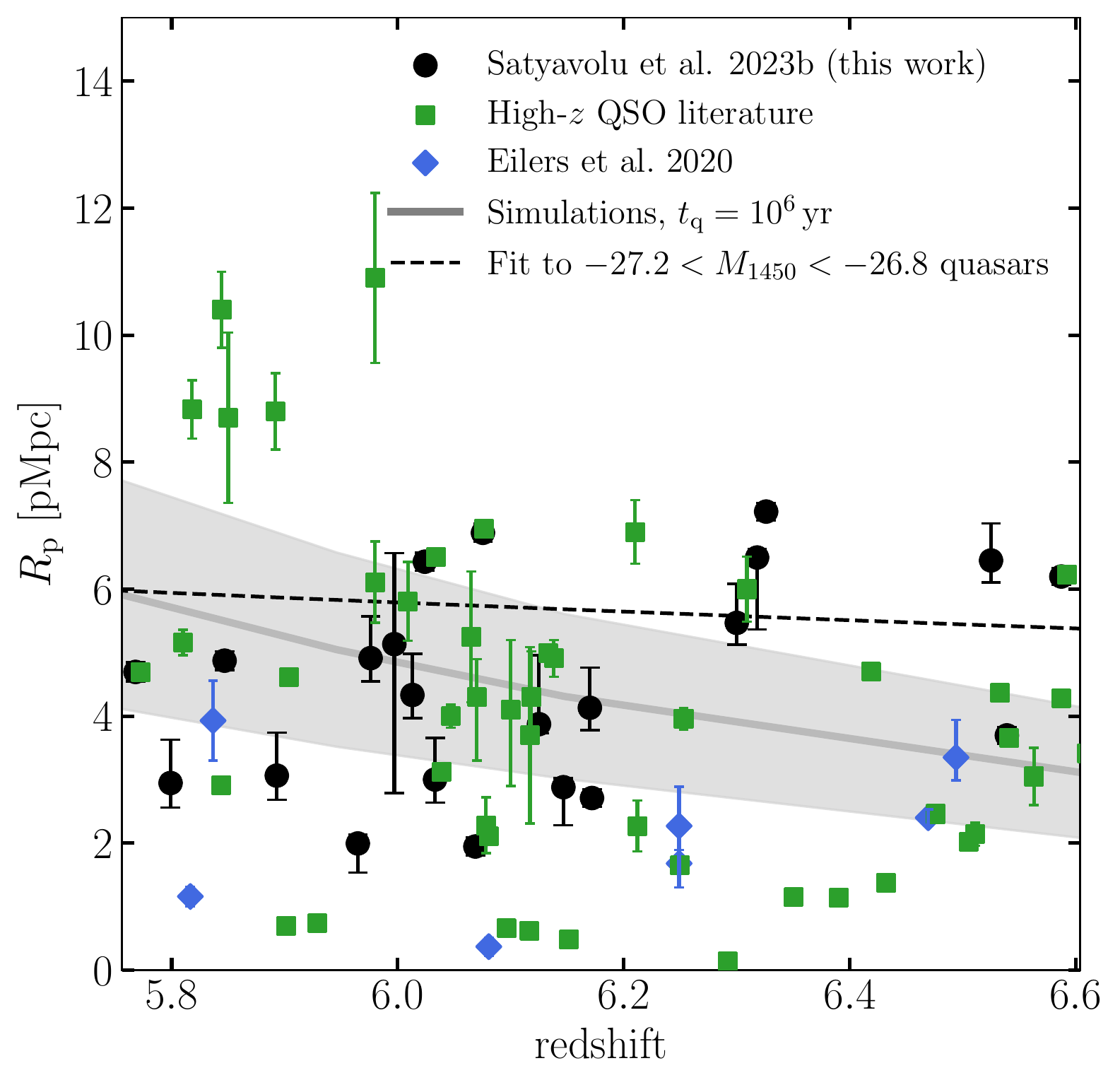}
  \caption{Evolution of proximity zone sizes. Older measurements are shown in green. The targeted sample of \citet{2020ApJ...900...37E} is shown in blue. Our measurements are shown in black. Also shown are the simulated proximity zones for a quasar of magnitude $-27$ and age of 1~Myr across different redshifts. The shaded region shows 68\% scatter across 500 sightlines in our simulation. The black dotted line shows the best fit curve $R_{\mathrm{p}}\propto (1+z)^{-0.89}$ to our measurements and previous measurements excluding \citet{2020ApJ...900...37E}.  For obtaining the best fit, only a relatively homogeneous subset of quasars, with $-26.8<M_{1450}<-27.2$ was used. \label{fig:rpvzdat}}
\end{figure}

Figure~\ref{fig:rpvzdat} shows the proximity zone sizes from all measurements including those presented in this paper. In order to study the evolution of proximity zone size with redshift, we fit to all measured proximity zone sizes  \citep[excluding the targeted sample of][] {2020ApJ...900...37E} including ours for a relatively homogeneous subsample of quasars with magnitudes between $-26.8$ and $-27.2$, assuming a power law between $\rp$ and  $(1+z)$. Unlike previous analyses, we do not correct the proximity zones to a common luminosity to get a best-fit. This is because the scaling between proximity zone sizes and magnitude is strongly dependent on the redshift of the quasar, and the same scaling cannot be applied to all quasars. Moreover, different measurements use a different scaling to obtain the luminosity-corrected proximity zones, which makes them unsuitable for comparison. 

We find a very shallow trend of $\rp\propto(1+z)^{-0.89}$, shallower than the previous inferences that were made through the luminosity-scaled proximity zones. This trend suggests that the scatter in the proximity zone sizes for similar magnitude quasars, as seen in Figure~\ref{fig:rpvmagfit}, is more likely due to the differences in their lifetimes. Indeed, one can notice that two of the farthest quasars with $z>6.5$ have larger than average proximity zone sizes, with an average luminosity. Although the universe is more neutral at higher redshifts, such large proximity zones can be explained by longer quasar lifetimes.  Smaller proximity zones are in fact found close to the smallest redshifts in the sample, which could have suggested either large scatter in the ionization state between sightline to sightline or smaller quasar lifetimes, although the latter seems to be favoured by our simulations.

\subsection{Correlation of proximity zone sizes with black hole mass}

\begin{figure}
  \includegraphics[width=\columnwidth]{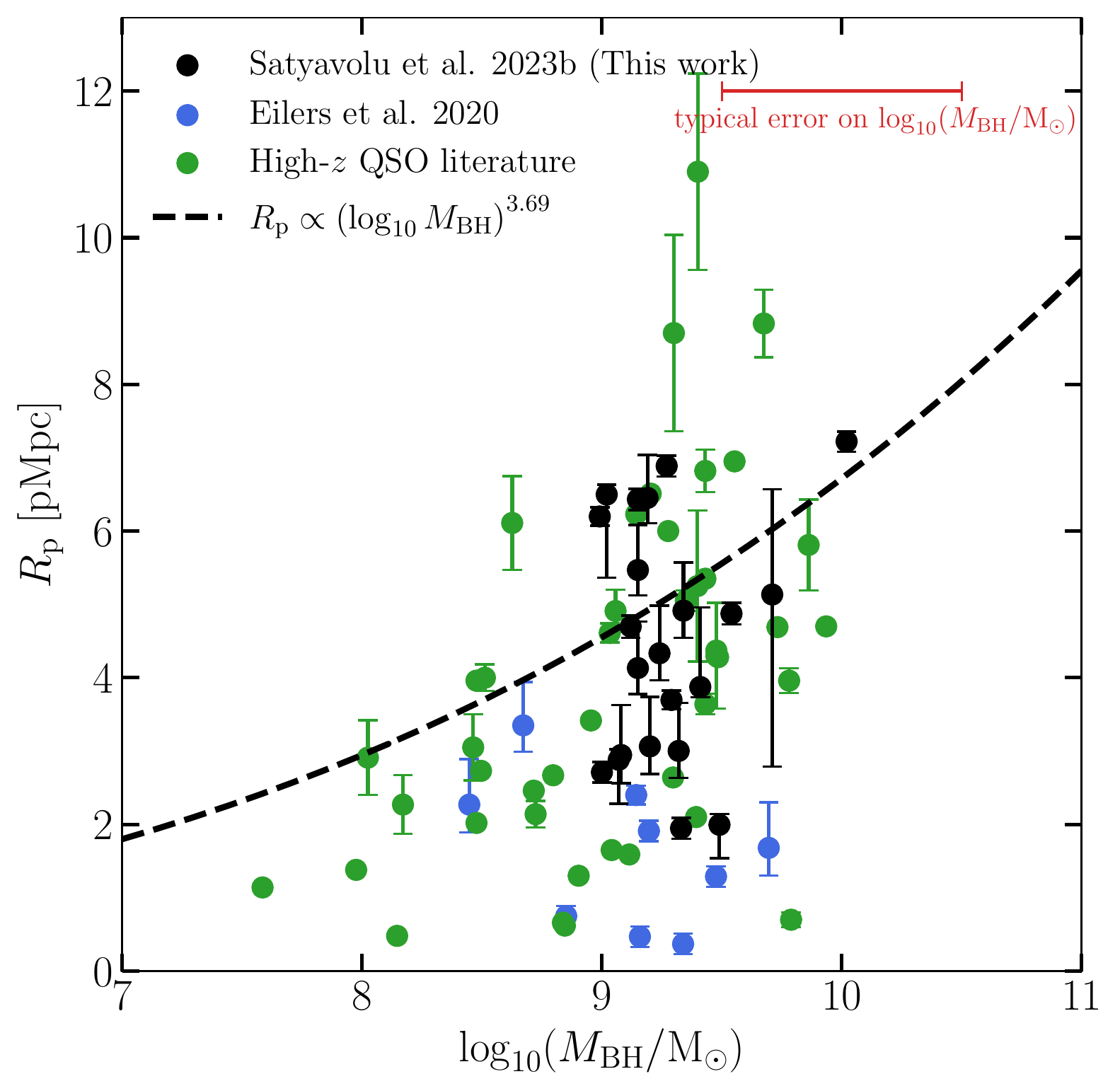}
  \caption{Proximity zone sizes as a function of supermassive black hole mass.  Previous measurements for which black hole masses were available are shown in green and the targeted sample from  \citet{2020ApJ...900...37E} is shown in blue. Our measurements are shown in black. Our black hole masses are from Mazzucchelli et al.\ (in preparation). The typical error on the black hole masses is represented by the errorbar at the top right in red. All black hole masses are based on \MgII\ linewidths. The black dotted line shows the best fit curve to our measurements and previous measurements excluding \citet{2020ApJ...900...37E}.  For obtaining the best fit, a relatively homogeneous subset of quasars with $-26.8<M_{1450}<-27.2$ and $6.0<z<6.2$ was used. A power-law relationship was assumed between the quasar proximity zone size and logarithm of the black hole mass, as motivated in the text. \label{fig:mbhvrp}}
\end{figure}

Proximity zone sizes are sensitive to the quasar lifetime \citep{2017ApJ...840...24E,2020MNRAS.493.1330D,2021ApJ...917...38E,2021ApJ...921...88M}.  As a result, combining proximity zone sizes with black hole mass measurements can potentially constrain the growth history of black holes \citep{2023MNRAS.521.3108S}. With this in mind, Figure~\ref{fig:mbhvrp} shows the proximity zone sizes of quasars in our sample against the masses of their central SMBHs.  The black hole masses for XQR-30 quasars were measured by Mazzucchelli et al. (in preparation), based on the Mg~II and C~IV linewidths, which can be used to derive the velocity of the gas clouds in the broad-line region and thereby the dynamical mass of the black hole, otherwise called the single-epoch viral black hole mass.  The black hole masses have a typical total uncertainty of 1 dex \citep{2009ApJ...699..800V}. The black hole masses of our sample are of the order $\sim 10^9\msun$, consistent with the other measurements at this redshift for comparable UV magnitudes \citep{2019ApJ...873...35S,2021ApJ...923..262Y,2022ApJ...941..106F}.  

For comparable UV magnitudes and redshifts, we expect the proximity zone sizes to increase with quasar lifetime, as in Equation~(\ref{eq:rion}). In an exponential growth model for the supermassive black hole, the black hole mass $M_{\mathrm{BH}}$ would be proportional to $\exp({\tq})$. We therefore try to fit a power law relation between the proximity zone size $\rp$ and the logarithm of the black hole mass, $\log_{10}M_\mathrm{BH}$, for a relatively homogeneous subset of quasars, with magnitudes $-26.8<M_{1450}<-27.2$ and redshifts $6<z<6.2$. We find a strong correlation of the proximity zone size with the black hole mass as $\rp\propto\mathrm{log_{10}}(M_\mathrm{BH})^{3.69}$, stronger than what is expected from Equation~(\ref{eq:rion}), which is valid only for lifetimes less than the equilibration timescale. This correlation is also stronger than what was inferred by \cite{2020ApJ...903...60I}. We plan to look for black hole growth models that are consistent with both the proximity zone sizes and black hole masses using simulations in a future work (Satyavolu et al., in preparation). 

\subsection{Correlation of proximity zones with closeness to metal absorption systems}

\begin{figure}
 \includegraphics[width=\columnwidth]{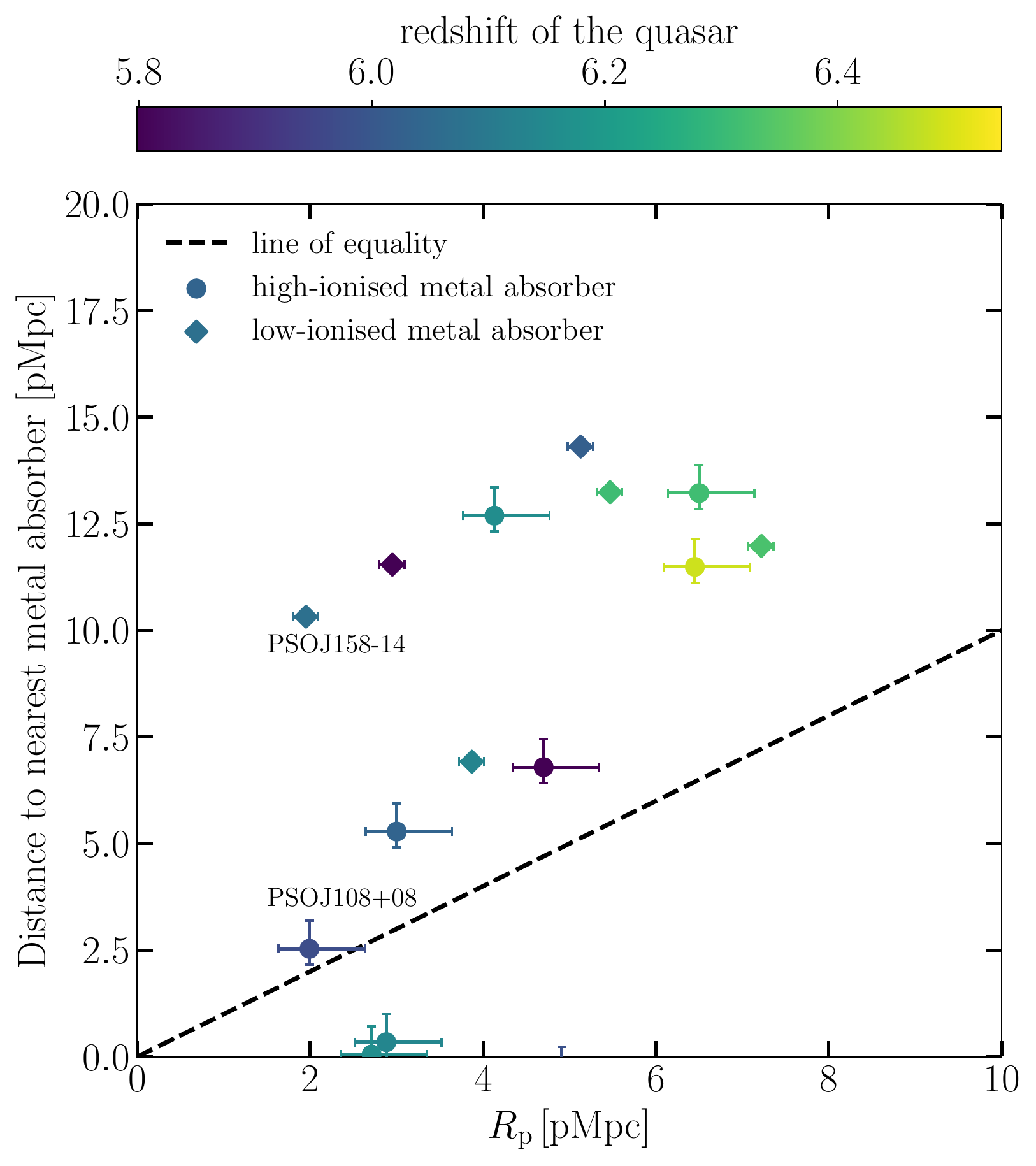}
  \caption{Distance to the nearest metal absorber as a function of proximity zone size. High-ionised metal absorbers are shown as circles while low-ionised metal absorbers are shown as diamonds. Colours represent the redshift of the metal absorber.  PSOJ108+08 is the only quasar in our sample with a metal-line absorber close to the edge of the proximity zone.  We also highlight PSOJ158--14 on this figure; this quasar is discussed in greater detail in Figure~\ref{fig:dwingprops}.}
  \label{fig:metalabs}
\end{figure}

Figure~\ref{fig:metalabs} shows the quasars in our sample for which the distance to the nearest metal absorber is within 20~pMpc.  Highly ionized absorbers are shown as circles while low-ionized systems are shown as diamonds.  Quasars with pDLAs and BALs are excluded from this sample. The ionised absorbers were identified by looking for absorption in additional transition lines corresponding to each ion through an automated search and visual inspection \citep{2023MNRAS.521..289D}. It can be seen that high and low ionization absorbers are found at both high and low redshifts in our sample. 

We find that quasar proximity zones fall into three categories. At the bottom of the plot, there are two quasars with relatively small proximity zones (2--3~pMpc) that house high ionization metal absorbers. For most quasars as seen in the top half of the plot, the closest metal absorption system sits beyond 10~pMpc from the quasar, well outside proximity zone boundary. For the quasars in our sample, there appears to be a strong correlation between proximity zone size and the presence of metal absorbers.  This could potentially be an effect of the quasar's ionizing radiation on the metal-line chemistry around it.  Low ionization metal absorbers, which may have more potential to truncate proximate zones, are found to cover the whole range of proximity zone sizes from 2 to 7~pMpc.  There are three proximity zones from 2 to 5~pMpc whose quasar lines of sight contain metal absorbers just outside the boundary of their proximity zones at a distance of 2.5 to 7~pMpc.  Only one quasar, PSOJ108+08, contains a metal absorber right at the edge of the proximity zone.  The lifetime of this quasar could be potentially underestimated as the proximity zone appears to be prematurely truncated. 

\subsection{Anomalously small proximity zones}\label{sec:smallrp}

Two quasars in our sample show particularly small proximity zones, with $\rp<2$ pMpc. These quasars are also at the brighter magnitude and lower redshift end of the range spanned by our sample, making it hard to explain the small proximity zone sizes without invoking a young quasar age.  While we leave a deeper investigation of these proximity zones for future work (Satyavolu et al., in preparation), we make some preliminary remarks here.

\subsubsection{PSOJ158-14}

The quasar PSOJ158-14 is at a redshift of 6.0687 with a magnitude of $-27.32$. The proximity zone size of this quasar is 1.95~pMpc.  \citet{2020ApJ...900...37E} have investigated this quasar and reported that it has a large star formation rate ($\sim 1420 \msun\,\mathrm{yr}^{-1}$),  large bolometric luminosity ($\sim 10^{47}$~erg~$\mathrm{s}^{-1}$), high Eddington ratio ($\lambda_{\mathrm{edd}}\sim$ 1), and shows signs of strong internal motions within the broad line region. They also point out the dust continuum emission of this quasar is very strong ($F_{\mathrm{cont}}\sim$ 3.46~mJy).

\begin{figure}
  \includegraphics[width=\columnwidth]{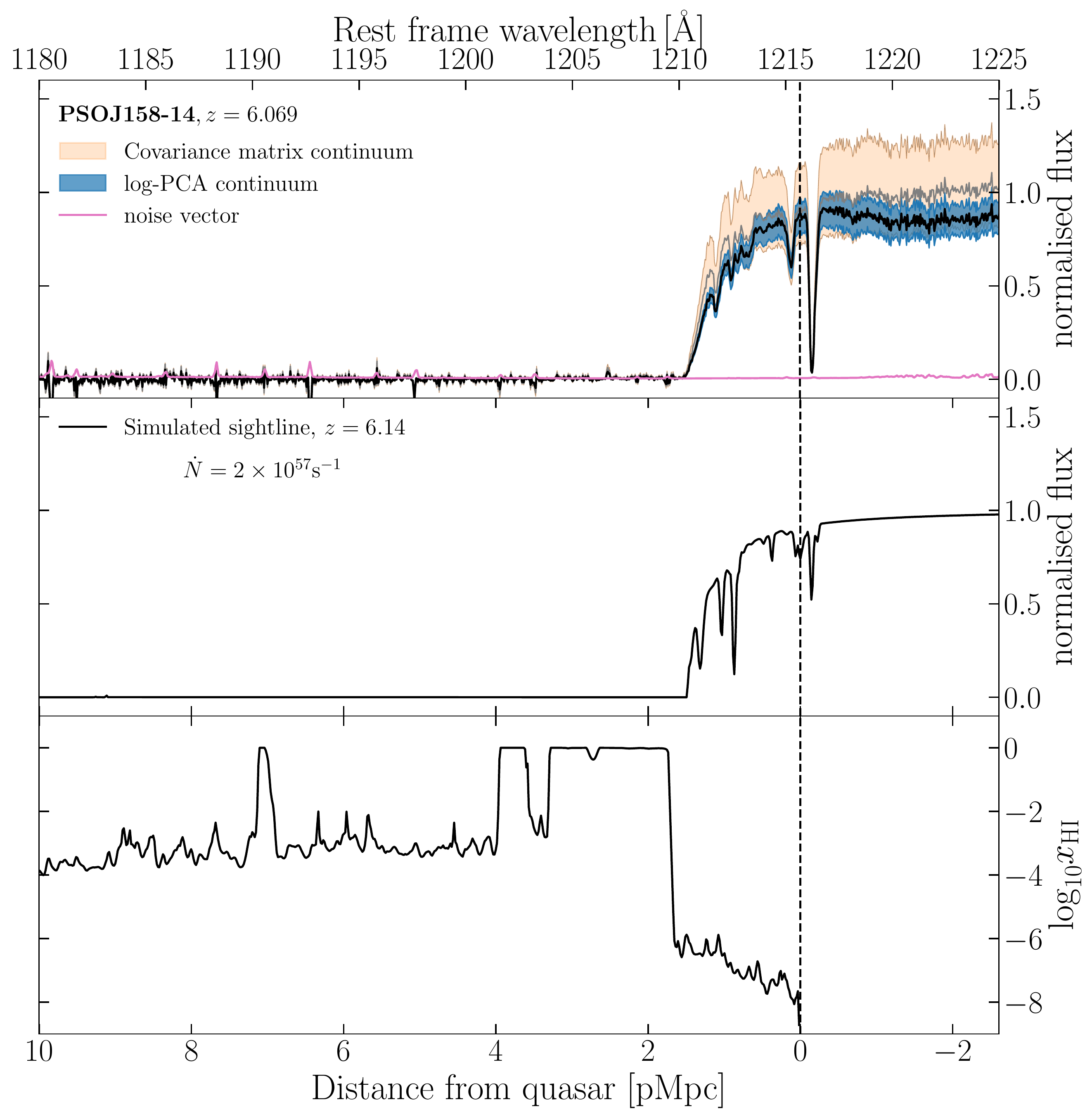}
  \caption{\textit{Top panel}: Continuum-normalised spectrum of PSOJ158-14, for two continuum reconstruction methods, the log-PCA method \citep[]{2022ApJ...931...29C} and the covariance matrix method \citep{2017MNRAS.466.1814G}, shown in blue and orange, respectively. Shaded regions show the $1\sigma$ spread around the median value.  (We use the log-PCA method for all quasars in this work.) \textit{Middle panel}: A simulated spectrum showing an IGM damping wing at $z=6.14$  for a quasar with magnitude $-27$ and age 1~Myr. \textit{Bottom panel}: The ionised hydrogen fraction along the same simulated sightline. This reveals the neutral hydrogen regions that create the damping wing seen in the middle panel. At redshift 6.14, only one of 500 sightlines in our simulation shows this feature.}
  \label{fig:dwingprops}
\end{figure}

Figure~\ref{fig:dwingprops} shows the continuum normalised spectrum of this quasar close to its \lya\ line.  We see that the spectrum blueward of the \lya\ line resembles a damping wing.  Additionally, the flux redward of the \lya\ line shows attenuation, as one would expect in the presence of a damping wing.  The flux continues to remain attenuated till $1233$~\AA.  Interestingly, there is no evidence of a compact high-column-density absorber ahead of the quasar.  The nearest metal absorber is at a redshift of 5.89874 \citep{2023MNRAS.521..289D}, which is well outside the edge of the proximity zone (at $\sim$ 10~pMpc from the quasar; see Figure~\ref{fig:metalabs}).  This suggests that if the feature around the \lya\ line of PSOJ158-14 is indeed a damping wing, it is likely to be caused by a neutral hydrogen `island' in the IGM.  Indeed, we do find similar sightlines in our simulation for comparable redshift and quasar brightness.  An example for $z=6.14$ and $M_{1450}=-27$ is shown in the middle panel of Figure~\ref{fig:dwingprops}.  This simulated sightline has a clearly visible damping wing, caused by a large neutral hydrogen patch in the IGM, which can be seen in the bottom panel of Figure~\ref{fig:dwingprops}.  For a quasar age of 1~Myr, only 1 of 500 simulated sightlines shows a damping wing.  For larger quasar lifetimes, this incidence drops.  For a quasar lifetime of 10~Myr, none of the simulated sightlines show a damping wing.  For a flickering lightcurve quasar, this number could be larger \citep{2023MNRAS.521.3108S}.

However, an IGM damping wing interpretation of the spectrum of PSOJ158-14 is less than certain.  Several aspects of this spectrum complicate its analysis.  For example, Figure~\ref{fig:dwingprops} also shows the continuum normalised flux for this quasar for a different continuum reconstruction, based on the covariance matrix method of \citet{2017MNRAS.466.1814G}.  We see that with this continuum, although the spectral shape still resembles a damping wing, the flux redward of the \lya\ does not appear to be attenuated.  Furthermore, when compared to the noise vector shown in Figure~\ref{fig:dwingprops}, the spectrum of PSOJ158-14 reveals flux just blueward of the proximity zone, suggesting that the damping-wing-like absorption might be not caused by the IGM.  While the evidence for this extended flux is relatively weak, the spectrum appears to have a statistically significant spike in flux at around $\sim 2$~pMpc from the edge of the proximity zone.  These considerations suggest that perhaps the spectrum is a result of absorption by a metal-poor absorber instead of the IGM.  In this scenario, the proximity zone size could be the result of a small quasar lifetime of $<10^4$~yr, and the flux bluewards of the proximity zone could be explained by residual flux from partial covering of the quasar continuum, or weak \lya\ emission from the absorber.  

More data seem to be necessary to rule out an IGM damping wing for this quasar. But if confirmed, PSOJ158-14 would be an interesting exception to the finding by \citet{2022arXiv221206907F} that a quasar with both small proximity zone and damping wing has not be found below redshift 7 so far.  

\subsubsection{PSOJ108+08}

The quasar PSOJ108+08 is at a relatively lower redshift of 5.9647 with a magnitude of $-27.59$. This quasar has the second smallest proximity zone size in our sample, with $R_\mathrm{p}=1.99$~pMpc. As we see in Figure~\ref{fig:fullsample}, the spectrum of this quasar does not show a damping wing. Although the proximity zone size is small, the flux blueward of the \lya\ line extends all the way up to $\sim 6$~pMpc i.e., nearly 3 times the proximity zone size and immediately increases above our 10\% threshold beyond the proximity zone.  We find a high-ionization metal absorber at 2.53~pMpc from this quasar, indicating that the proximity zone might be prematurely truncated due to absorption of the quasar flux by this absorber.  

PSOJ108+08 suggests that to better estimate lifetimes in such quasars, it might be worthwhile to explore alternate definitions for the proximity zone, such as defining the proximity zones as points where the flux transmission is at 5\% as well as 20\% and changing the smoothing length, which we will explore in future work (Satyavolu et al.\ in preparation).

\section{Conclusions}
\label{sec:conclusions}

We measured proximity zone sizes of 22 quasars at redshifts between 5.8 and 6.5 and UV magnitudes $M_{1450}$ between $-26$ and $-29$ using high-SNR spectra obtained with the X-SHOOTER instrument on the VLT telescope. Of the 22 quasar spectra that we study, 14 were obtained as part of the XQR-30 survey. The other eight quasars were obtained with X-SHOOTER from previous programs and were chosen to have similar resolution and SNR to the XQR-30 spectra. We summarize our results below:

\begin{itemize}
\item The proximity zone sizes of our quasars range from 1.95 to 7.22~pMpc.  This roughly corresponds to quasar lifetimes of $10^4$ to $10^8$~yr in the lightbulb model.  About 9\% of our measured proximity zones are small, requiring lifetimes of less than $10^4$~yr.  This distribution of proximity zone sizes is consistent with previous measurements of quasars with similar magnitudes and redshifts.  This work increases the number of available proximity zone size measurements at $z > 5.7$ to 87.
\item We update the proximity zone size measurements of 10 quasars previously studied in the literature, with the help of updated spectra and redshifts. The new measurements are consistent with previous measurements within 1--5\%.
\item We infer a scaling of proximity zone size with UV magnitude based on all measurements for quasars within the redshift range $6<z<6.2$ and find it to be consistent with our expectations from simulations. This scaling is shallower than what was measured previously \citep{2020ApJ...903...60I}.
\item We infer a scaling of proximity zone size with redshift based on all measurements for quasars with magnitudes $-27.2<M_{1450}<-26.8$ and find it to be shallower than what was measured from previous analyses \citep{2017ApJ...840...24E,2017ApJ...849...91M,2020ApJ...903...60I}. The shallowness of this scaling suggests that the scatter in the proximity zone sizes for quasars of similar UV magnitudes is a result of variation in quasar lifetimes.
\item We infer a scaling of proximity zone size with black hole mass and find it to be steeper than what is expected from theory.  Previously, \citet{2020ApJ...903...60I}  reported little to no correlation between $\rp$ and black hole mass.
\item Two of our quasars have exceptionally small $\rp$ of less than 2~pMpc.  One of these quasars shows possible signatures of a damping wing produced by the intergalactic medium or an extremely metal-poor foreground galaxy. Another has a high-ionized metal absorber close to the edge of the proximity zone.  

\end{itemize}

Our measurements of proximity zone sizes, and their correlations with quasar brightness, redshift, and black hole mass point towards a diverse range of quasar lifetimes.  The overall picture remains consistent with our previous finding that proximity zone size measurements seem to support a scenario in which supermasive black holes at high redshifts undergo obscured growth \citep{2023MNRAS.521.3108S}.  In a follow-up paper (Satyavolu et al., in preparation), we plan to discuss the quasar lifetime estimates based on the proximity zone size distribution measured in this work, which will lead to constraints on obscuration fractions, black hole seed masses, and black hole seed redshifts. 

\section*{acknowledgments}

GK is partly supported by the Department of Atomic Energy (Government of India) research project with Project Identification Number RTI~4002, and by the Max Planck Society through a Max Planck Partner Group.  This work was supported by grants from the Swiss National Supercomputing Centre (CSCS) under project IDs s949 and s1114.  This work is based on observations collected at the European Southern Observatory under ESO large program 1103.A-0817(A). SEIB and FW acknowledge funding from the European Research Council (ERC) under the European Union's Horizon 2020 research and innovation programme (grant agreement no. 740246 ``Cosmic Gas''). EPF\ is supported by the international Gemini Observatory, a program of NSF's NOIRLab, which is managed by the Association of Universities for Research in Astronomy (AURA) under a cooperative agreement with the National Science Foundation, on behalf of the Gemini partnership of Argentina, Brazil, Canada, Chile, the Republic of Korea, and the United States of America. This research was supported by the Australian Research Council Centre of Excellence for All Sky Astrophysics in 3 Dimensions (ASTRO 3D), through project number CE170100013. Support by ERC Advanced Grant 320596 'The Emergence of Structure During the Epoch of reionization' is gratefully acknowledged. MGH acknowledges the support of the UK Science and Technology Facilities Council (STFC) and the National Science Foundation under Grant No. NSF PHY-1748958.  For the purpose of open access, the author has applied a Creative Commons Attribution (CC BY) licence to any Author Accepted Manuscript version arising from this submission. This paper is based on the following ESO observing programmes: XSHOOTER 60.A-9024, 084.A-0360, 084.A-0390, 084.A-0550, 085.A-0299, 086.A-0162, 086.A-0574, 087.A-0607, 088.A-0897, 091.C-0934, 294.A-5031, 096.A-0095, 096.A-0418, 097.B-1070, 098.B-0537, 0100.A-0625, 0101.B-0272, 0102.A-0154, 0102.A- 0478, 1103.A-0817; HAWKI (H band data) 105.20GY.005. 

\section*{Data availability}
The data and code underlying this article will be shared on reasonable request to the corresponding author. The reduced spectra will be released with the XQR-30 survey paper \citep{xqr30general}.

\bibliographystyle{mnras}
\bibliography{xqr30_refs} 

\appendix

\section{Quasar Spectrum}

Figure~\ref{fig:fullsample_cont} shows all quasar spectra analysed in this paper. 

\begin{figure*}
  \begin{center}
    \includegraphics[scale=0.75]{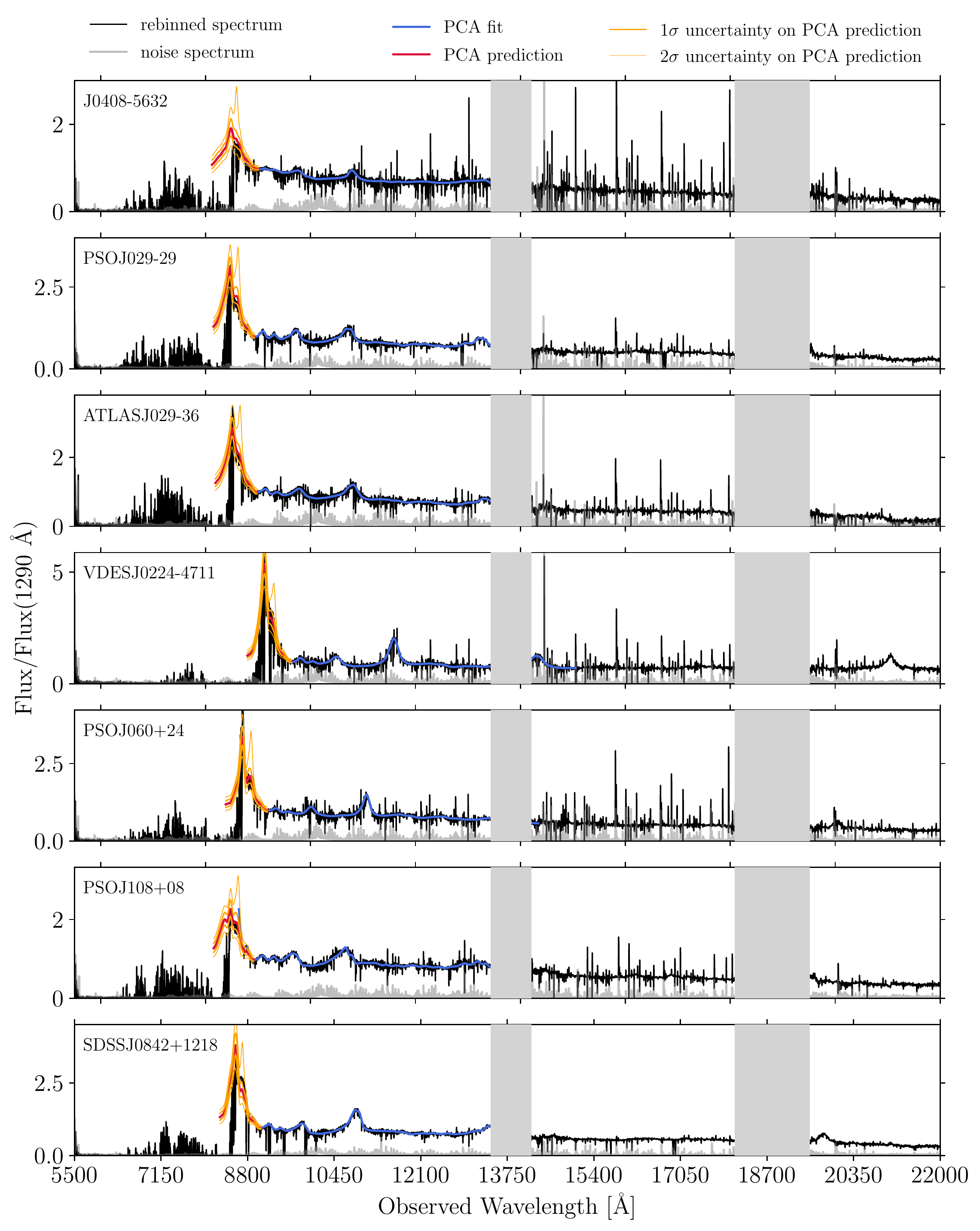}
    \caption{All spectra studied in this paper. Rescaled flux is obtained by dividing with the flux at 1290 \AA. This rescaled flux, rebinned to 50 km/s, is shown in black. Uncertainty on the flux is shown in grey. Telluric absorption bands are shown in light grey. Best fit continuum to the redside flux is shown in blue. Predicted blue-side continuum is shown in red. 1 and 2 $\sigma$ errors on the continuum predictions are shown in orange.\label{fig:fullsample_cont}}
  \end{center}
\end{figure*}

\begin{figure*}
  \begin{center}
    \includegraphics[scale=0.75]{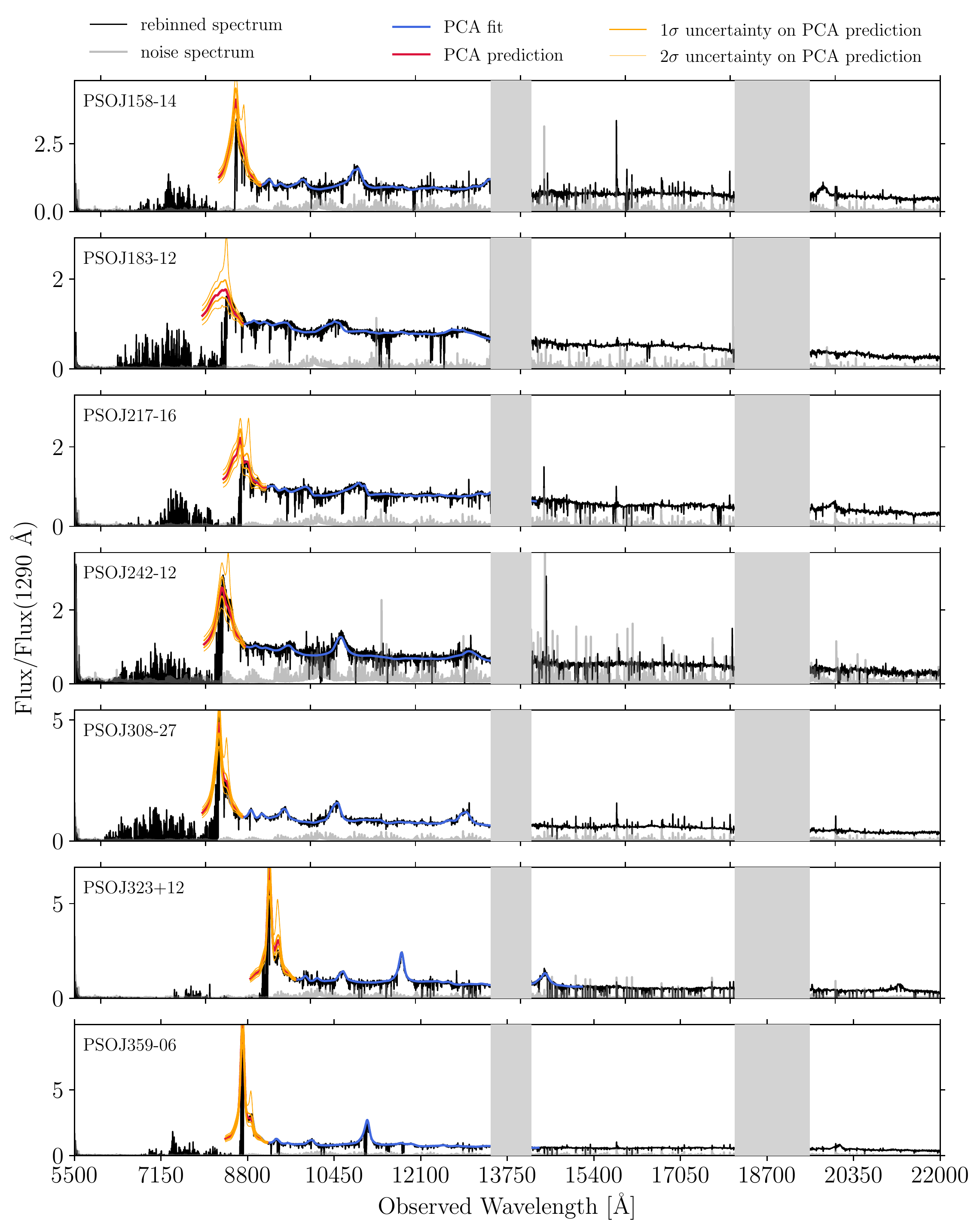}
    \contcaption{}
  \end{center}
\end{figure*}

\begin{figure*}
  \begin{center}
    \includegraphics[scale=0.75]{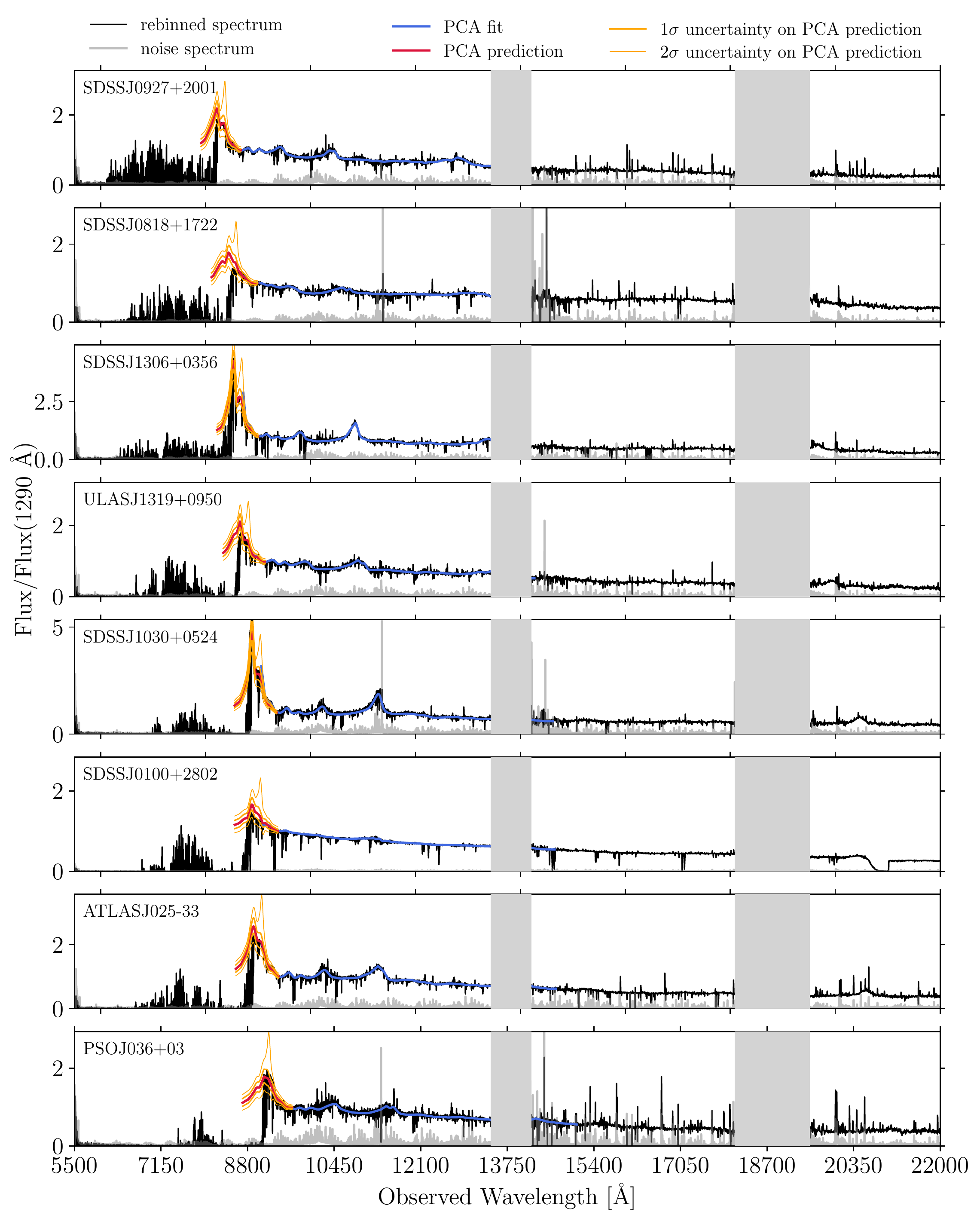}
    \contcaption{}
  \end{center}
\end{figure*}

\bsp
\label{lastpage}
\end{document}